\documentclass[sigconf]{acmart}

\usepackage{enumitem}
\usepackage{url}

\usepackage{algorithm}

\usepackage{makecell}  % 셀 안 줄바꿈용

\usepackage{xcolor}

\newcommand{\readText}[1]{\textcolor[HTML]{0070C0}{#1}}

\newcommand{\matchText}[1]{\textcolor[HTML]{C00000}{#1}}

\definecolor{LightRed}{RGB}{255,200,200}
\definecolor{LightBlue}{RGB}{200,220,255}

\newif\ifshowComment

\newif\ifshowComment

\newif\ifshowComment

\newif\ifshowComment

\newif\ifshowComment

\newif\ifshowComment

\newif\ifshowComment

\colorlet{strongorange}{red!90!black!90}

\newcommand{\CHG}[1]{#1}

\showCommentfalse  % false일 때 주석 표시 X

\AtBeginDocument{%
  }

\setcopyright{acmlicensed}
\copyrightyear{2026}
\acmYear{2026}
\setcopyright{cc}
\setcctype{by-nc-nd}
\acmConference[CHI '26]{Proceedings of the 2026 CHI Conference on Human Factors in Computing Systems}{April 13--17, 2026}{Barcelona, Spain}
\acmBooktitle{Proceedings of the 2026 CHI Conference on Human Factors in Computing Systems (CHI '26), April 13--17, 2026, Barcelona, Spain}
\acmPrice{}
\acmDOI{10.1145/3772318.3791455}
\acmISBN{979-8-4007-2278-3/2026/04}

\begin{document}

\title{Draped Surfaces: A Contour-\CHG{Adaptive} Interface Overlaid on the Physical Environment for Mixed Reality Workspaces}

\author{SoonUk Kwon}

\affiliation{%
  \institution{University of British Columbia, Okanagan}
  \city{Kelowna}
  \state{BC}
  \country{Canada}
}
\email{kwonars@student.ubc.ca}

\author{Barrett Ens}

\affiliation{
  \institution{University of British Columbia, Okanagan}
  \city{Kelowna}
  \state{BC}
  \country{Canada}}
\email{barrett.ens@ubc.ca}

\author{Pourang Irani}

\affiliation{
  \institution{University of British Columbia, Okanagan}
  \city{Kelowna}
  \state{BC}
  \country{Canada}
}
\email{pourang.irani@ubc.ca}

\renewcommand{\shortauthors}{Kwon et al.}

\begin{abstract}

Conventional Mixed Reality (MR) workspaces are frequently organized in cockpit-like layouts, where multiple floating windows surround the user. While this configuration facilitates access to digital content, it often induces occlusion, reducing understanding of the physical environment and limiting access to real-world objects. To overcome this challenge, we present the Contour-Adaptive Mixed Environment Overlays (CAMEO), a contour-adaptive MR interface that \textit{drapes} virtual windows onto physical surfaces. This design integrates digital content with nearby items, thereby improving users’ visual access to background objects and supporting interaction with them. We evaluate CAMEO in two controlled studies. The first demonstrates that draping reduces hand-movement detours relative to flat mid-air surfaces, enabling more direct interaction with nearby items. The second shows that controlled window deformation does not significantly impair text legibility when compared to flat surfaces. Together, these findings contribute a novel design paradigm for MR workspaces that balances immersion, readability, and environmental understanding.

\end{abstract}

\begin{CCSXML}
<ccs2012>
   <concept>
       <concept_id>10003120.10003121.10003124.10010392</concept_id>
       <concept_desc>Human-centered computing~Mixed / augmented reality</concept_desc>
       <concept_significance>500</concept_significance>
       </concept>
   <concept>
       <concept_id>10003120.10003121.10003124.10010866</concept_id>
       <concept_desc>Human-centered computing~Virtual reality</concept_desc>
       <concept_significance>500</concept_significance>
       </concept>
 </ccs2012>
\end{CCSXML}

\ccsdesc[500]{Human-centered computing~Mixed / augmented reality}
\ccsdesc[500]{Human-centered computing~Virtual reality}

\keywords{Text Legibility, Visual Clutter, Shape Perception, Spatial Awareness, Environmental Awareness, Contour-Adaptive Overlays, Motor Behavior (Motor Behaviour), Occlusion, Mixed Reality (MR), Augmented Reality (AR), Virtual Reality (VR), Spatial Computing, Diffusion-Based Mesh Smoothing}

\begin{teaserfigure}
  \includegraphics[width=\textwidth]{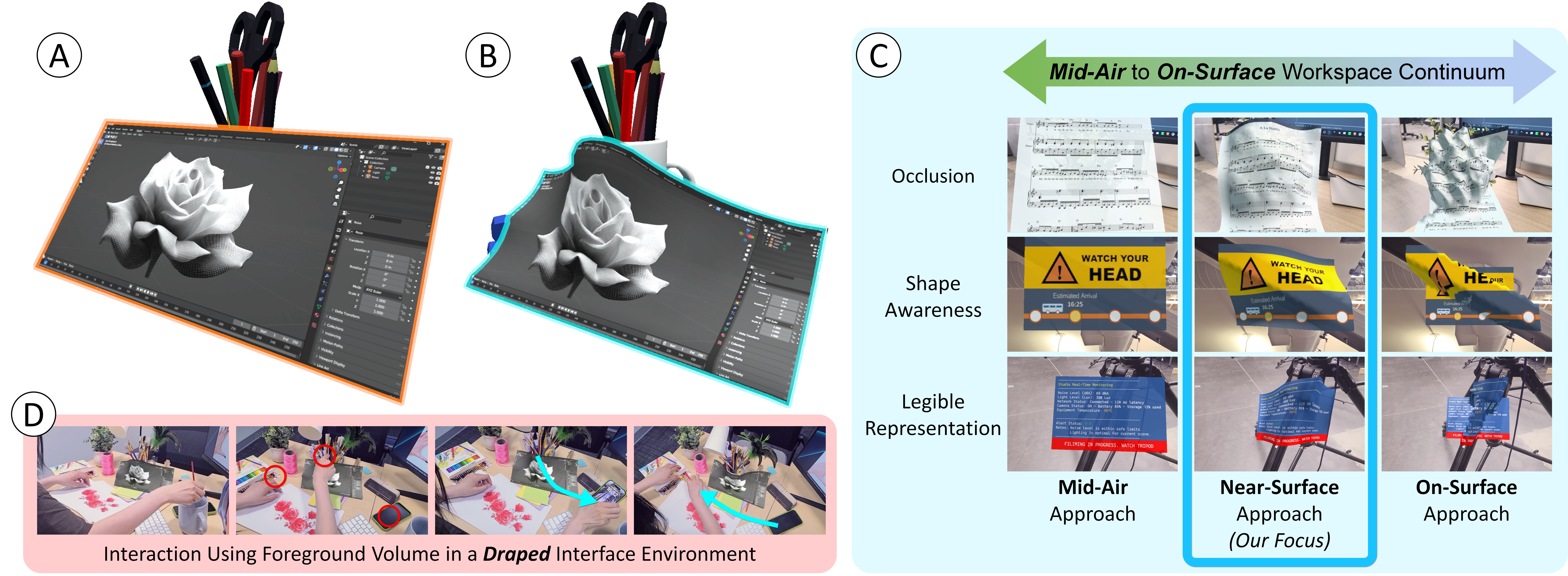}
  \caption{
    A: The flat interface is represented. B: The interface is being “\textit{draped,}” onto a nearby object, addressing the trade-off between legibility and background occlusion. C: Mid-Air to On-Surface workspace continuum. In the Mid-Air region, the interface struggles to conform to the environment, which hinders immediate environment understanding and reduces placement flexibility. Conversely, as the workspace approaches the On-Surface region, the interface becomes less legible due to complex geometries. In the middle, our approach balances legibility with environmental understanding. D: An example interface setting using our Near-Surface approach. The red circles indicate the user’s interaction points around the interface, while the cyan traces represent hand movements. By placing the interface close to physical objects, users gain more foreground volume (in front of the interface) and improved view of their surroundings.}
  \Description{The figure illustrates how interface behavior changes across the continuum between on-surface and mid-air workspaces, highlighting differences in legibility, environmental understanding, and interaction in mixed reality. Panel A presents a flat interface used as a baseline. Panel B shows the interface being draped, illustrating how draping balances the trade-off between legibility and immediate environment understanding. Panel C shows the Mid-Air to On-Surface continuum. When the interface is directly attached to a physical surface in the On-Surface region, its legibility decreases because it must conform tightly to complex object geometries. In contrast, when positioned in the Mid-Air region, the interface no longer conforms to the environment, making immediate environment understanding more difficult and reducing placement flexibility. 
  Panel D shows an example of surrounding objects with which the user interacts, such as a colored pencil, a paint jar, and a smartphone, illustrating how the interface is used in this type of real-world environment. By positioning the interface close to these physical objects, the workspace provides more usable foreground volume and less occluded area, allowing users to maintain better awareness of their surroundings and ultimately supporting smoother and more natural interactions in mixed-reality environments.
}
  \label{fig:teaser}
\end{teaserfigure}

\maketitle

\section{Introduction} \label{sec:introduction}

Mixed Reality (MR) mid-air windows, arranged in three-dimensional (3D) space much like traditional multi-monitor setups~\cite{cheng2023interactionadapt, cheng2021semanticadapt, li2019holodoc} (Fig.~\ref{fig:teaser}(A)), extend desktop displays by offering increased workspace capacity, flexible spatial organization, and new opportunities for multitasking~\cite{ens2014personal, ens2014ethereal}. 
However, \CHG{such layouts can reduce users’ understanding of background objects and overall scene clarity~\cite{cheng2023interactionadapt, cheng2021semanticadapt, li2024situationadapt, lindlbauer2019context}, which in turn limits their ability to interact with nearby objects~\cite{lommertzen2008collision, regan2000visually}.}
Alternative arrangements, including semi-transparent MR interfaces can improve environmental understanding, but result in limited performance~\cite{hussain2024effect} due to issues of depth perception~\cite{kruijff2010perceptual} and transparency~\cite{macedo2021occlusion, hussain2024effect}. Despite recent efforts, including semi-transparent MR interfaces~\cite{hussain2024effect, gabbard2006effects, gattullo2015legibility, di2015text, rzayev2018reading} and \CHG{intelligent layout methods~\cite{lu2022exploring}, sustaining a functional awareness of objects} in MR workspaces remains a critical yet insufficiently addressed challenge.

Existing approaches to extending workspaces into 3D environments have sought to more closely integrate digital content with the user’s physical surroundings. Spatially Augmented Reality (SAR) systems, for example, project interfaces onto walls, tables, or other surfaces~\cite{rekimoto1999augmented, raskar1999spatially}, or distribute content throughout entire rooms~\cite{kim2023perspective}. While promising, these systems struggle with complex geometries that can fragment or distort projected content~\cite{ichihashi2019estimating} \CHG{(Fig.~\ref{fig:teaser}(C) right)}. Head-mounted display (HMD) systems, by contrast, typically employ floating virtual windows~\cite{feiner1993windows, li2019holodoc, cools2022towards, pavanatto2024multiple, ens2014personal, ens2016shared} (Fig.~\ref{fig:teaser}(C) left) or immersive environments~\cite{zhou2022depth, berki2019desktop} to expand digital workspaces. Although effective at increasing capacity and flexibility, \CHG{these designs often prioritize immersion over environmental integration, leaving users with diminished understanding of surrounding objects and spatial context}. Parallel work in computer graphics has addressed smoothing 3D shapes~\cite{tukey1974nonlinear, taubin1995signal, tomasi1998bilateral, chen2023geometric} and mapping textures according to designer intent~\cite{sun2013texture}. However, these advances have rarely been considered from the perspective of interface usability. Similarly, research on tangible user interfaces (TUI) has demonstrated ways to leverage physical objects as interactive surfaces~\cite{faleel2021hpui, harrison2011omnitouch, du2022opportunistic, he2024adaptui, xiao2013worldkit}, but such approaches have not enabled the seamless extension of existing user interface (UI) structures across general physical environments. This gap underscores the need for workspace designs that expand digital interaction while preserving continuity with the physical world.

Prior work on legibility in SAR~\cite{di2015text}, HMD-based AR~\cite{gattullo2015legibility}, and curved surfaces~\cite{wei2020reading, park2017effects, jeong2017consumers} has largely focused on uniform geometries, leaving the challenge of maintaining legibility on complex, irregular shapes underexplored. To address these limitations, we introduce the \CHG{Contour-Adaptive Mixed Environment Overlays (CAMEO)}, a deformable workspace that drapes MR windows over the user's real-world surroundings. 
Drawing on insights from prior occlusion research~\cite{medeiros2022shielding}, our approach extends the 2D planar MR workspace into an object contour-adaptive interface (Fig.~\ref{fig:teaser}(B)). Unlike conventional flat MR workspaces, \CHG{CAMEO} is designed to simultaneously preserve both, the legibility of digital content and enable users to perceive and interact with objects in their immediate environment (Fig.~\ref{fig:teaser}(D)). Positioned between flat and fully deformable paradigms (Fig.~\ref{fig:teaser}(C) middle) \CHG{CAMEO} introduces a novel design space for MR work environments that aims to mitigate \CHG{occlusion of background objects and support interaction with them.} 

We assessed \CHG{CAMEO} in a 3-part simulation and found that draped UIs occlude less background and secure more foreground space in comparison to flat MR interfaces. We also observed that smoothing iterations reveal \textit{Balanced Points} between view and space efficiency as the surface converges toward planarity.
In addition, \CHG{our simulation study reveals the regions for full visibility of the UI while preserving shape information.}
Through two user studies, \CHG{we found that CAMEO reduces the number of hand detours required to interact with the user's nearby object in comparison to flat mid-air workspaces interfaces.} \CHG{We also found that across diverse draped UI conditions, CAMEO supported reading performance that did not differ significantly from that of flat interfaces.} Finally, we observed that due to its draped contours, \CHG{CAMEO} facilitates the inference of objects behind the interface, an ability not afforded by conventional flat designs. \CHG{This ability to maintain gentle geometric variations that reveal background regions while still conveying underlying shapes and content through subtle depth and contour cues has not been previously demonstrated.}

Our key contributions are as follows:
\begin{itemize}
    \item A novel contour-adaptive workspace, \CHG{CAMEO, that drapes MR windows over surrounding real-world geometry while maintaining interface legibility};
    \item Empirical evidence that \CHG{CAMEO} enhances users’ ability to access and interact with nearby physical objects;
    \item \CHG{Designs opportunities} for novel, deformable workspace layouts and interactions enabled by \CHG{CAMEO}.
\end{itemize}

\section{Related Works}\label{sec:relatedworks}

Research on workspace design has progressively expanded from traditional 2D desktops to large displays, multi-monitor setups, and immersive MR environments. Building on this trajectory, we review prior work across four key themes: workspace expansion, legibility on complex geometries, environmental understanding, and deformable MR workspaces.

\subsection{Workspace Expansion Through Mixed Reality}

Traditional graphical user interfaces (Graphical UIs or GUIs) have historically focused on organizing digital content within bounded visual spaces such as desktop windows \cite{wadlow1981xerox, johnson1989xerox}. As the constraints of limited screen real estate became evident, researchers explored various methods to optimize window management and layout strategies \cite{greenberg1986issues, funke1993approach}. While many of these efforts emphasized expanding the display space via large screens \cite{bi2009comparing, ni2006increased, endert2012designing, czerwinski2003toward}, multi-monitor setups \cite{cetin2018visual, wallace2014effect, cauchard2011visual}, or through HMD-based ethereal MR displays \cite{ens2014ethereal, li2019holodoc, pavanatto2024multiple}, a more recent shift has moved beyond merely enlarging screens to contextually integrating interfaces within the physical environment \cite{cheng2023interactionadapt, cheng2021semanticadapt, li2024situationadapt, lindlbauer2019context}. These attempts to expand into MR have achieved advances in integrating with the environment by utilizing dynamic placement \cite{li2024situationadapt} and optimization \cite{cheng2023interactionadapt, cheng2021semanticadapt} of the interface considering the environment, but there remain inherent limitations in accessing environmental information behind such layouts.
Early work, such as Augmented Surfaces~\cite{rekimoto1999augmented}, contributed to expanding digital workspaces across existing physical surfaces. Building on this foundation, subsequent research explored integration using projection-based systems~\cite{waldner2011display, riemann2018flowput, cotting2006interactive} and spatially augmented-reality (SAR) methods~\cite{raskar1998office, raskar1999spatially, kim2023perspective}. Such projector-based approaches present clear environmental information, but they either support only specific types of geometries~\cite{gronbaek2020kirigamitable, follmer2013inform} or have limitations in rendering interfaces on all discontinuous surfaces where the shape changes abruptly, such as edge regions~\cite{cotting2006interactive, riemann2018flowput, kim2023perspective, ichihashi2019estimating}.
Although MR interfaces have become increasingly integrated with the physical world, most still fail to fully account for complex surface geometries, which are critical for both environmental understanding and legibility.

\subsection{Legibility Within Complex Geometric Shapes}

When virtual interfaces are embedded onto irregular or non-planar surfaces in the user's physical environment, legibility becomes a critical usability concern. Because UI content is directly affected by the underlying geometry of the physical surfaces~\cite{ichihashi2019estimating}, maintaining textual clarity is essential for effective and meaningful interaction in MR settings.
Several studies have investigated legibility by examining factors such as resolution~\cite{dittrich2013legibility}, lighting conditions~\cite{sousa2017vrrrroom, gabbard2006effects}, color contrast~\cite{gattullo2015legibility}, surface patterns~\cite{di2015text}, and text positioning~\cite{chen2004testbed, rzayev2018reading}. Although the research most relevant to ours focuses on surface shape and legibility~\cite{park2017effects, wei2020reading, jeong2017consumers, robertson1993document}, these studies have either examined legibility on displays composed of joined planar surfaces~\cite{park2017effects, robertson1993document} or focused on surfaces with uniform curvature~\cite{wei2020reading, jeong2017consumers}. While some studies raise concerns about focal distance shifts~\cite{tan2003effects, gabbard2018effects}, others suggest that non-planar interfaces may improve readability and reduce visual fatigue~\cite{shupp2009shaping}. However, legibility on arbitrarily curved surfaces remains largely unexplored, indicating a need for further investigation in this area.
Meanwhile, the field of computer graphics has extensively explored techniques for smoothing irregular geometries~\cite{taubin1995signal, tomasi1998bilateral, tukey1974nonlinear} and for texture mapping onto complex shapes~\cite{sun2013texture}. However, from the standpoint of legible interface design, little research has investigated how users read text or recognize symbols when they are presented on arbitrarily curved physical surfaces. Addressing this gap in legibility on arbitrarily curved surfaces, a central focus of our work, will make it possible to extend desktop MR layouts that may mitigate \CHG{occlusion.}

\subsection{Environmental Understanding in Mixed Reality Interfaces}

To make effective use of physical environments, interface design must go beyond simple legibility and incorporate visual cues from physical objects to support real-world interaction~\cite{sivak1990integration, regan2000visually, lommertzen2008collision}.
However, MR interfaces that are overlaid onto the physical environment without considering its contours, can impair depth perception and reduce users’ understanding of their surroundings~\cite{kruijff2010perceptual, macedo2021occlusion, phillips2020veiled, davari2020occlusion}. Flat MR interfaces that account for physical geometry may work in simple settings. However, they often fail in complex 3D environments, where effective placement for spatial understanding becomes significantly more difficult. Transparent interfaces provide background visibility by blending with the physical environment. However, such approaches can cause visual clutter~\cite{hussain2024effect, satkowski2021investigating, macedo2021occlusion}, especially when users view flat interfaces head-on to maximize the screen’s field of view~\cite{ens2014personal}.
Reducing the interface size may help minimize obstruction, but the extent of reduction is constrained by the limited amount of information that can be presented~\cite{diverdi2004level, lindlbauer2019context}. Users may also manually reposition or remove interfaces, but this does not offer a sustainable solution~\cite{lu2022exploring}. Taken together, these trade-offs indicate the need to further explore alternative approaches. 

\subsection{Compatible Deformable Mixed Reality Workspaces}

More recent research began to explore diverse physical contexts, including tangibility achieved by leveraging physical environments \cite{faleel2021hpui, follmer2013inform, gronbaek2020kirigamitable, du2022opportunistic, he2024adaptui, xiao2013worldkit, henderson2009opportunistic}. While these approaches reveal the potential to utilize previously overlooked environmental geometric features, they fall short in enabling the compatible reuse of UI layouts originally designed for existing applications. This limitation becomes particularly evident when interfaces must adapt to arbitrary physical environments, such as those in users' vicinity.
By visually draping existing interface components as-is, over real-world surfaces, our method expands the workspace into the proximate environment, which enables diverse interaction possibilities~\cite{cheng2022comfortable, lischke2016screen}.
Ultimately, our method provides a foundation for deformable MR workspaces that can adapt fluidly to physical surroundings, allowing digital interfaces to inhabit the ``ether~\cite{ens2014ethereal}'' in a way that respects the complexity of real-world contexts. Building on these prior works (Table \ref{tab:related_works}), we position \CHG{CAMEO} within the broader landscape of MR workspace layout management.

\begin{table*}[th]
\centering
\begin{tabular}{@{} p{3cm} c c c c c @{}}
\toprule
\makecell[c]{\textbf{Previous}\\ \textbf{Work}} &
\makecell[c]{\textbf{Display}\\ \textbf{Type}} &
\makecell[c]{\textbf{Shape}\\ \textbf{Awareness}} &
\makecell[c]{\textbf{Environment}\\ \textbf{Agnostic}} &
\makecell[c]{\textbf{Legible}\\ \textbf{Representation}} &
\makecell[c]{\textbf{UI Layout}\\ \textbf{Conservation}} \\
\midrule
KirigamiTable~\cite{gronbaek2020kirigamitable}    & Projector-based & $\textcolor{green!90!black!65!blue}\bigcirc$ & $-$ & $\textcolor{orange!80!yellow}\triangle$ & $\textcolor{green!90!black!65!blue}\bigcirc$ \\
\CHG{inFORM}~\cite{follmer2013inform}                   & Projector-based & $\textcolor{green!90!black!65!blue}\bigcirc$ & $\textcolor{orange!80!yellow}\triangle$ & $\textcolor{orange!80!yellow}\triangle$ & $\textcolor{green!90!black!65!blue}\bigcirc$ \\
FlowPut~\cite{riemann2018flowput}                 & Projector-based & $\textcolor{green!90!black!65!blue}\bigcirc$ & $\textcolor{green!90!black!65!blue}\bigcirc$ & $\textcolor{orange!80!yellow}\triangle$ & $\textcolor{green!90!black!65!blue}\bigcirc$ \\
\CHG{Display Bubbles}~\cite{cotting2006interactive}     & Projector-based & $\textcolor{green!90!black!65!blue}\bigcirc$ & $\textcolor{green!90!black!65!blue}\bigcirc$ & $\textcolor{orange!80!yellow}\triangle$ & $\textcolor{green!90!black!65!blue}\bigcirc$ \\
SituationAdapt~\cite{li2024situationadapt}        & HMD-based       & $\textcolor{orange!80!yellow}\triangle$ & $\textcolor{green!90!black!65!blue}\bigcirc$ & $\textcolor{green!90!black!65!blue}\bigcirc$ & $\textcolor{green!90!black!65!blue}\bigcirc$ \\
InteractionAdapt~\cite{cheng2023interactionadapt} & HMD-based       & $\textcolor{orange!80!yellow}\triangle$ & $\textcolor{green!90!black!65!blue}\bigcirc$ & $\textcolor{green!90!black!65!blue}\bigcirc$ & $\textcolor{green!90!black!65!blue}\bigcirc$ \\
\CHG{Sublimate}~\cite{leithinger2013sublimate}          & HMD-based       & $\textcolor{green!90!black!65!blue}\bigcirc$ & $\textcolor{orange!80!yellow}\triangle$ & $\textcolor{green!90!black!65!blue}\bigcirc$ & $\textcolor{green!90!black!65!blue}\bigcirc$ \\
BlendMR~\cite{han2023blendmr}                     & HMD-based       & $\textcolor{green!90!black!65!blue}\bigcirc$ & $\textcolor{green!90!black!65!blue}\bigcirc$ & $\textcolor{green!90!black!65!blue}\bigcirc$ & $-$ \\
CAMEO (Ours)                                      & HMD-based       & $\textcolor{green!90!black!65!blue}\bigcirc$ & $\textcolor{green!90!black!65!blue}\bigcirc$ & $\textcolor{green!90!black!65!blue}\bigcirc$ & $\textcolor{green!90!black!65!blue}\bigcirc$ \\
\bottomrule
\end{tabular}
\caption{Comparison of previous works by display type and interface characteristics: Shape Awareness, Environment Agnostic, Legible Representation, and UI Layout Conservation. ($\textcolor{green!90!black!65!blue}\bigcirc$: Supported, $\textcolor{orange!80!yellow}\triangle$: Partially Supported, $-$: Not Supported)}
\label{tab:related_works}
\end{table*}

\section{Contour-Adaptive Mixed Environment Overlays: Design Principles and Implementation}\label{sec:CAMEO}

We first describe the design principles guiding our implementation of \CHG{CAMEO} followed by implementation details. We address the interface placement challenge from a user-centered perspective, drawing on prior work showing that spherical arrangements of rectilinear screens~\cite{ens2014personal} can enhance visibility from the user’s viewpoint. To further support background understanding, \CHG{CAMEO} \textit{drapes} the MR interface over the user's environment. However, fully affixing the interface to the environment's surface geometry (as in projected-based display type~\cite{riemann2018flowput} or SAR~\cite{kim2023perspective}) can reduce legibility, while rectangular floating overlays can limit understanding of the environment. We focus on this inherent trade-off. To address this issue, \CHG{we consolidated insights from existing occlusion literature~\cite{medeiros2022shielding} and our exploratory workshop (see Appendix A) into two sets of \CHG{CAMEO} Design Requirements (CDRs)}: CDR1: \textit{Interface Legibility} and CDR2: \textit{\CHG{Environmental} Understanding}.

\begin{itemize}[label={}]
\item\textbf{CDR1: Interface Legibility}
    \begin{enumerate}
        \item \label{item:iv1} \textit{Every part of the interface must remain visible from the user’s viewpoint;}
        \item \label{item:iv2} \textit{Each part of the interface has as much as attainable, a uniform size and orientation from the user's perspective.}
    \end{enumerate}
\item\textbf{CDR2: \CHG{Environmental} Understanding}
    \begin{enumerate}
        \item \label{item:eu1} \textit{Every part of the interface follows as closely as possible the contours of the background occluded environment;}
        \item \label{item:eu2} \textit{The region occupied by the interface remains as minimal as possible.}
    \end{enumerate}
\end{itemize}

These requirements are established to make the interface visible from the user's viewpoint, convey background information, and minimize the area the interface occupies in the user's field of view. Our \CHG{CAMEO} implementation fulfills these requirements through two stages: initial shape formation and surface smoothing.

\begin{figure}[htbp]
    \centering
    \includegraphics[width=\columnwidth]{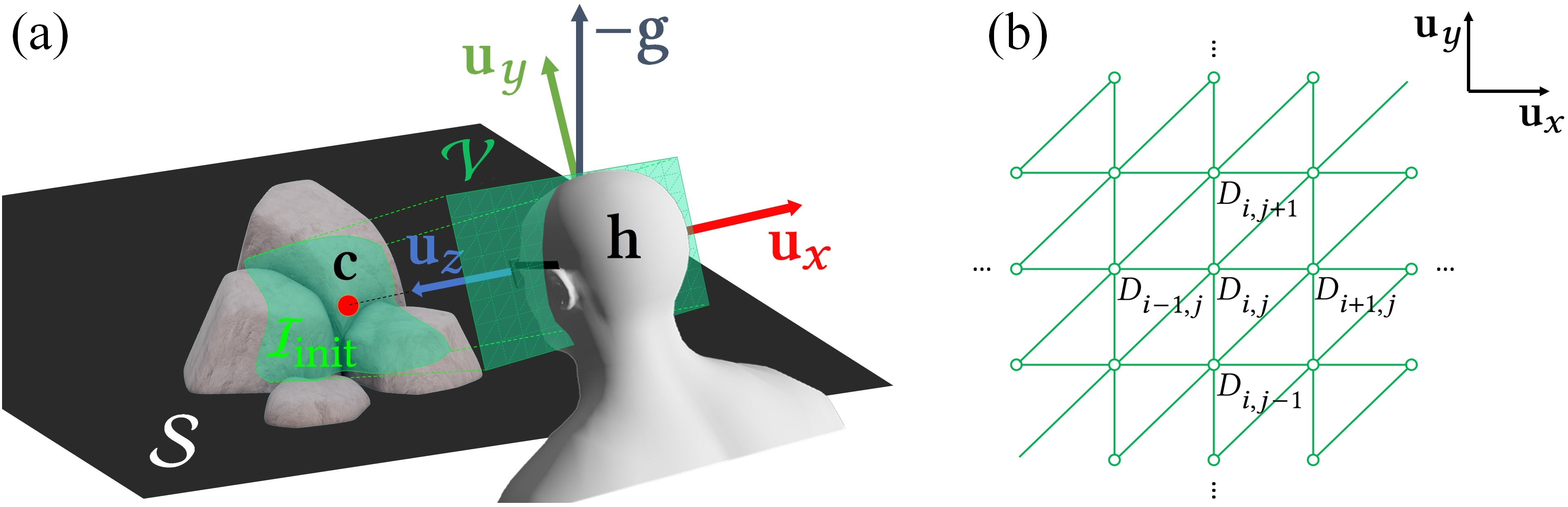}
\caption{(a) Initialization of the initial interface $\mathcal{I}_\text{init}$. 
\CHG{The red dot shows the interface center $\mathbf{c}$ arbitrarily set by the user, while $\mathbf{h}$ and $-\mathbf{g}$ indicate the user’s viewpoint and the opposite gravity direction. The forward direction $\mathbf{u}_z$ is defined using $\mathbf{h}$ and $\mathbf{c}$, setting the interface placement line. The upward direction $\mathbf{u}_y$ is obtained by projecting $-\mathbf{g}$ onto the plane normal to $\mathbf{u}_z$, and the rightward direction $\mathbf{u}_x$ is orthogonal to both. The green grid represents the ray origins $\mathcal{V}$, from which rays along $\mathbf{u}_z$ intersect the surface $\mathcal{S}$ to form $\mathcal{I}_\text{init}$. We empirically set the grid resolution to 1 cm in all experiments to sufficiently capture the geometry of the objects.}
(b) Topology of the interface and indexing scheme, where each $D_\text{(indices)}$ represents the depth value from $\mathcal{V}$ to $\mathcal{S}$. The interface is constructed as these triangles form a mesh. For more details, refer to Appendix~B.2.}
\Description{This figure explains how the system initializes the interface and how the interface’s depth values are organized. Part (a) describes how the initial interface is positioned in 3D space. A single dot marks the center of the interface, which the user chooses freely. The user’s view direction is represented as a forward-pointing vector from the eyes, and gravity is represented as a downward direction; the figure also includes the opposite of gravity, which points upward. Using the line from the user’s viewpoint to the interface center, the system defines a forward axis (i.e., a placement line) for positioning the interface. It then defines an upward axis by projecting the opposite-gravity direction onto a plane perpendicular to the forward axis. A rightward axis is computed as the vector perpendicular to both the forward and upward axes. Together, these three axes define the orientation of the interface. A green rectangular grid is defined at the head center; these grid points role as the starting positions for rays cast forward toward the object surface to construct the interface. Wherever these rays intersect the surface, the intersection points define the shape of the initial interface. The spacing between grid points is set to one centimeter to capture the object’s shape with sufficient detail. Part (b) shows how the interface is represented as a mesh. From the intersection points described in Part (a), adjacent points are connected to form triangles, and all these triangles together create the full initial surface mesh of the interface. In addition, the distances between each ray’s origin and its intersection point, corresponding to ray lengths, are stored, which is later used in the interface smoothing process. More information is provided in Appendix B.2.}
    \label{fig:initialization}
\end{figure}

\subsection{Extracting Contour Information: Initial Condition of \CHG{CAMEO}}

\CHG{The first stage sets the interface’s position and orientation based on the user-centered perspective scheme \cite{ens2014personal}, placing it toward the user’s forward direction (Fig.~\ref{fig:initialization}). A rectangular array of ray origins $\mathcal{V}$ is placed on a plane orthogonal to $\mathbf{u}_z$ and centered at $\mathbf{h}$. Rays from $\mathcal{V}$ along $\mathbf{u}_z$ intersect the surface $\mathcal{S}$, and the intersection points define $\mathcal{I_\text{init}}$ (Eq.~\ref{eq:init}).}
\begin{equation}
\mathcal{I}_{\mathrm{init}} = \left\{\, \mathbf{I}^{\mathrm{init}}_{i,j} \;\middle|\; \mathbf{I}^{\mathrm{init}}_{i,j} = \mathbf{v}_{i,j} + D_{i,j}^{\text{init}} \mathbf{u}_z
\right\}
\label{eq:init}
\end{equation}

Here, $\mathbf{I}^{\mathrm{init}}_{i,j}$ denotes a vertex of $\mathcal{I}_{\mathrm{init}}$, $\mathbf{v}_{i,j}$ the ray's origin in $\mathcal{V}$, and $D_{i,j}^{\text{init}} \in \mathbb{R}$ denotes the initial depth value at index $(i,j)$ of the depth map. As such, the initial interface $\mathcal{I_\text{init}}$ now incorporates contour information from background objects.

\subsubsection{Infinite Ray Problem and Adaptive Maximum Depth $D_\text{max}$.}

\begin{figure*}[htbp]
    \centering
    \includegraphics[width=\textwidth]{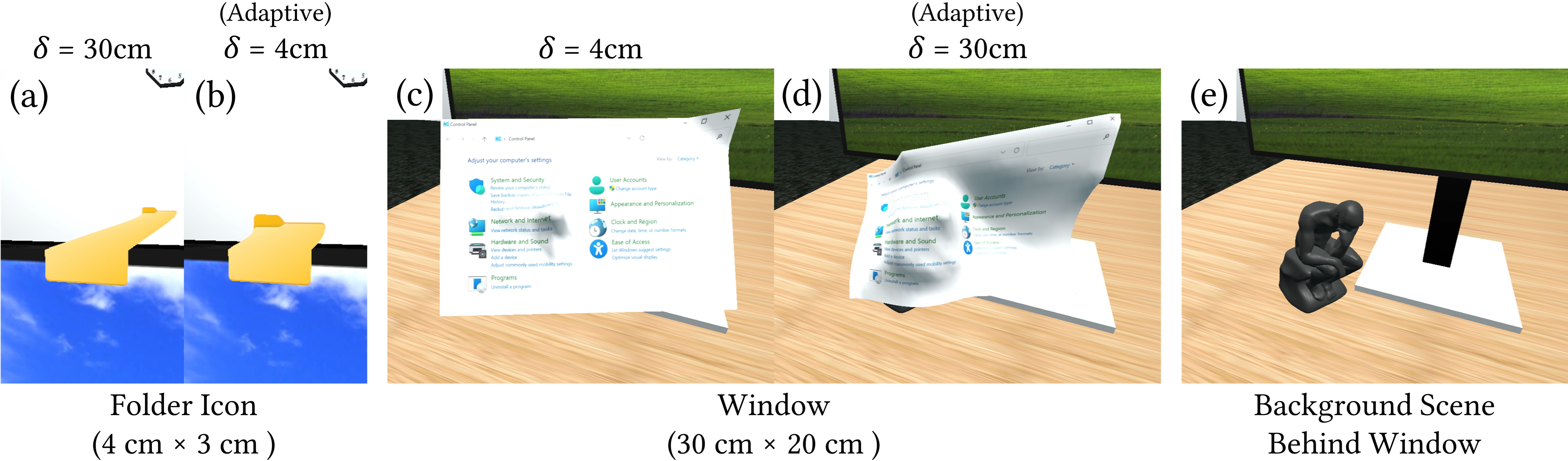}
    \caption{\CHG{Adaptive effect of $D_{\max}$ with respect to interface size, visualized by varying an arbitrary $\delta$ in $D_{\max} = D_{\min} + \delta$ as an illustrative example to aid the understanding of Eq.~\ref{eq:initial_max_depth}.}
    (a) A 4cm × 3cm folder icon placed excessively far when $\delta = 30$ cm, making the icon largely distorted. 
    (b) The icon using adaptive $D_{\max}$ in Eq.~\ref{eq:initial_max_depth}, making the icon less distorted.
    (c) A 30cm × 20cm window placed excessively close when $\delta = 4$ cm, making objects and empty space indistinguishable.
    (d) The window using adaptive $D_{\max}$ in Eq.~\ref{eq:initial_max_depth}, making objects and empty space distinguishable.
    (e) Background scene before window placement.
    Adaptive cases (b) and (d) are smoothed by the Balanced Point (Section~\ref{subsec:balancedpoint}). 
    Textures: Windows 11 Folder icon and Control Panel window © Microsoft.}
    \Description{This figure explains how the maximum depth range of an interface automatically adjusts depending on the size of the interface. To help illustrate this idea, the figure shows several examples where the depth range is intentionally set either too large or too small, and compares them with cases where the adaptive version of the depth range is used. Part (a) shows a small folder icon, about four by three centimeters in size. In this example, the allowed depth range is set far too large, causing the icon to be placed much farther away than it should be. As a result, the icon becomes heavily distorted and no longer appears as intended. Part (b) shows the same folder icon, but this time using the adaptive depth range defined by our method. With the adaptive setting, the icon appears at a more appropriate distance, reducing distortion and making the icon easier to recognize. Part (c) shows a large window, about thirty by twenty centimeters. Here, the allowed depth range is set much too small, causing the window to be placed too close to the viewer. At this close distance, the content inside the window becomes difficult to distinguish from the empty space around it. Part (d) shows the same large window using the adaptive depth range. With the adaptive setting, the window is placed at a proper viewing distance, making both the objects inside it and the empty regions clearly separable.  Part (e) shows the background scene before any window is placed, included for comparison.  In the adaptive examples shown in parts (b) and (d), the resulting interface is also smoothed using the Balanced Point method described earlier in the paper.  The folder icon and the Control Panel window shown in the figure use textures from Windows 11 and are copyrighted by Microsoft.}
    \label{fig:dmax}
\end{figure*}

If a ray does not intersect the physical surface and extends into empty (or infinite) space, the \CHG{CAMEO} interface cannot be defined. Excessively deep vertices may cause large depth distortions. To address this, we define $D_{i,j}^{\text{init}}$ and $D_{\max}$ as Eq.~\ref{eq:initial_max_depth}.
\begin{equation}
D_{i,j}^{\text{init}} = \begin{cases}
D_{i,j}^* \\
D_{\max}
\end{cases}, \quad D_{\text{max}} = D_{\min} + \max(W, H)
\label{eq:initial_max_depth}
\end{equation}
Here, $D_{i,j}^*$ denotes the depth of the ray to $\mathcal{S}$ with $D_{i,j}^* < D_{\max}$. The upper bound $D_{\max}$ is adaptively determined by the interface width $W$, height $H$, and the minimum depth $D_{\min}$ among all rays hitting the physical surface. This formulation mitigates depth distortions, places the interface near the surface to reduce occlusions, and prevents infinitely extending rays. Consequently, both $D_{i,j}^*$ and $D_{\max}$ contribute to satisfying the CDR1-(\ref{item:iv2}), CDR2-(\ref{item:eu1}), and CDR2-(\ref{item:eu2}) (Fig.~\ref{fig:dmax}).

% Boundary Condition
\subsection{Smoothing While Respecting Physical Boundaries: Boundary Condition of \CHG{CAMEO}}\label{subsec:boundary_condition}

\subsubsection{Laplacian-based Smoothing Using Diffusion Equation}\- We adopted a Laplacian-based depth map smoothing using the diffusion (heat) equation~\cite{desbrun1999implicit} to \textit{diffuse} the depth information across the interface. This method ensures smoothness across the entire interface (Eq. \ref{eq:diffusion_equation}) while preserving its overall shape (\CHG{see Appendix C.4; a comparison with other smoothing techniques is provided in Appendix B.1}). We define the depth map at smoothing iteration $\text{n}$ as $D_{i,j}^\text{n}$ with initial values $D_{i,j}^0 = D_{i,j}^{\text{init}}$, so the interface at smoothing iteration $\text{n}$ is $\mathcal{I}^\text{n}$. To implement the numerical method, the continuous diffusion equation is discretized via the finite difference method~\cite{richardson1911ix} (Eq.~\ref{eq:diffusion_equation}). To discretize the continuous form, we treat $t$ as an artificial time variable linked to iteration $\text{n}$, and $x$ and $y$ as spatial variables along $\mathbf{u}_x$ and $\mathbf{u}_y$ vectors. This discretization yields the second line of Eq.~\ref{eq:diffusion_equation}.
\begin{equation}
\begin{gathered} % aligned 대신 gathered 사용
\frac{\partial D}{\partial t} = c \left( \frac{\partial^2 D}{\partial x^2} + \frac{\partial^2 D}{\partial y^2} \right) = c\nabla^2 D \\ 
\Rightarrow D_{i,j}^\text{n+1} = \frac{1}{4} \left( D_{i+1,j}^\text{n} + D_{i,j+1}^\text{n} + D_{i-1,j}^\text{n} + D_{i,j-1}^\text{n} \right)
\end{gathered}
\label{eq:diffusion_equation}
\end{equation}
At interface edges where the update is not applicable, $D_{i,j}^{\text{n+1}}$ is set equal to the value from the line immediately before the edge. Further details of the smoothing process are provided in the Appendix B.3, B.4.

\subsubsection{Boundary Condition}
\begin{figure}[htbp]
    \centering
    \includegraphics[width=\linewidth]{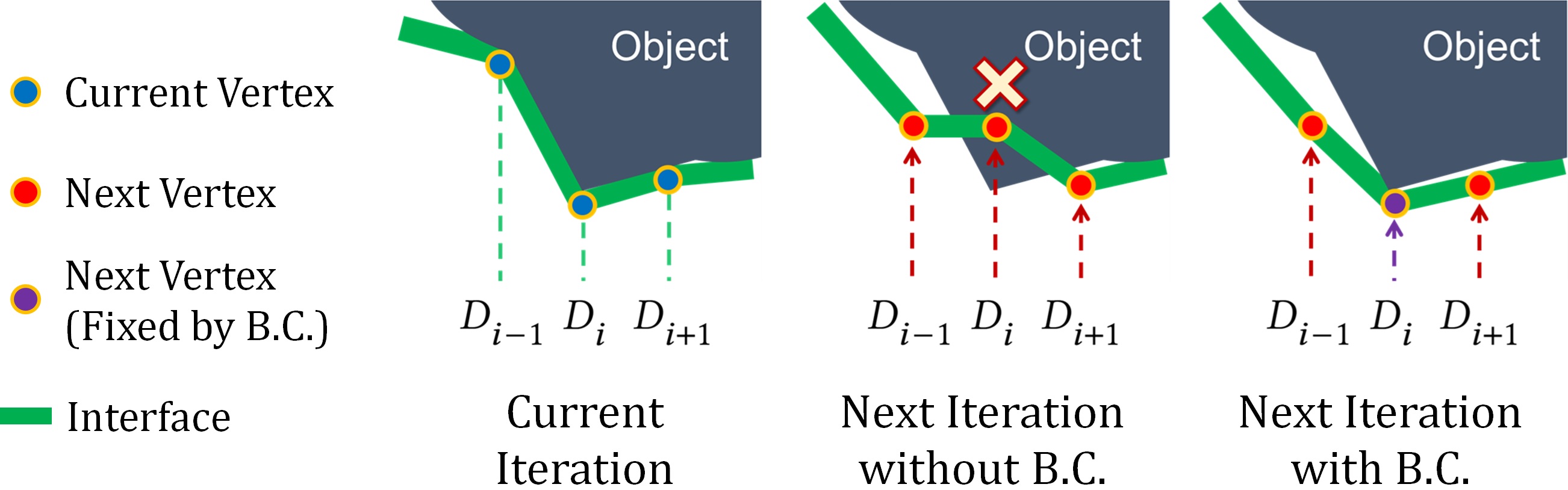}
    \caption{Operation of the boundary condition (B.C.). Blue circles denote the vertices at the current iteration, while red and purple circles represent those of the next iteration, and the green line indicates the interface. Without a boundary condition, the smoothing operation causes interface vertices to move inside the object, preventing an accurate representation of the object shape. With a boundary condition, vertices displaced into the interior are invalidated and revert to their initial positions, allowing the interface to correctly capture the geometry of the object. For clarity, this figure is simplified by considering only the case of index $i$.}
    \Description{This figure explains how the boundary condition works during the smoothing process. The blue circles represent the positions of the interface vertices during the current step of smoothing. The red and purple circles represent where those points would move in the next iteration, and the green line shows the interface shape. Without the boundary condition, the smoothing process can cause some points to move inward, resulting in the interface being placed inside the object. When this happens, the interface can no longer represent the true shape of the object. With the boundary condition applied, any point that moves inside the object is treated as invalid and is returned to its initial location. By doing this, the interface remains on the surface of the object, allowing it to accurately match the object’s shape.}
    \label{fig:boundary}
\end{figure}

We impose a boundary condition to prevent the surface from penetrating the object (Fig.~\ref{fig:boundary}). If
$
D_{i,j}^\text{init} < \frac{1}{4}\left(D_{i+1,j}^\text{n} + D_{i,j+1}^\text{n} + D_{i-1,j}^\text{n} + D_{i,j-1}^\text{n}\right),
$
then $D_{i,j}^{\text{n+1}}$ is reset to the initial depth $D_{i,j}^{\text{init}}$. This ensures that mesh vertices do not move deeper than their original positions during smoothing, preserving object geometry and maintaining a monotonic decrease in depth. For more details, see Appendix B.3.

\subsection{When Should We Stop Smoothing? Distortion vs. Occlusion}\label{subsec:balancedpoint}

Our approach is designed to preserve essential geometric information about the physical environment while maintaining a low distortion for interface legibility. However, these two objectives are inherently in conflict and, moreover, the need to consistently adapt interfaces to a broad range of surface geometries further underscores the difficulty of this challenge.
\CHG{To address this tension, we determine the maximum iteration number $\text{C}$ using a balanced formulation that explicitly trades off between two quantitative metrics designed to reflect our CDRs: View Efficiency $E_\text{view}$ and Space Efficiency $E_\text{space}$. These metrics are grounded in quantifiable geometric cues, ensuring consistent behavior regardless of where the interface is applied. The View Efficiency $E_\text{view}$ is derived from Screen Entropy $H_\text{screen}$, which represents the uniformity of interface mesh elements, while the Space Efficiency $E_\text{space}$ is derived from Available Foreground Volume $S_\text{avail}$, which represents the occluded space behind the interface. Each of the definitions is presented in the following sections.}
\begin{figure*}[htbp]
    \centering
    \includegraphics[width=\textwidth]{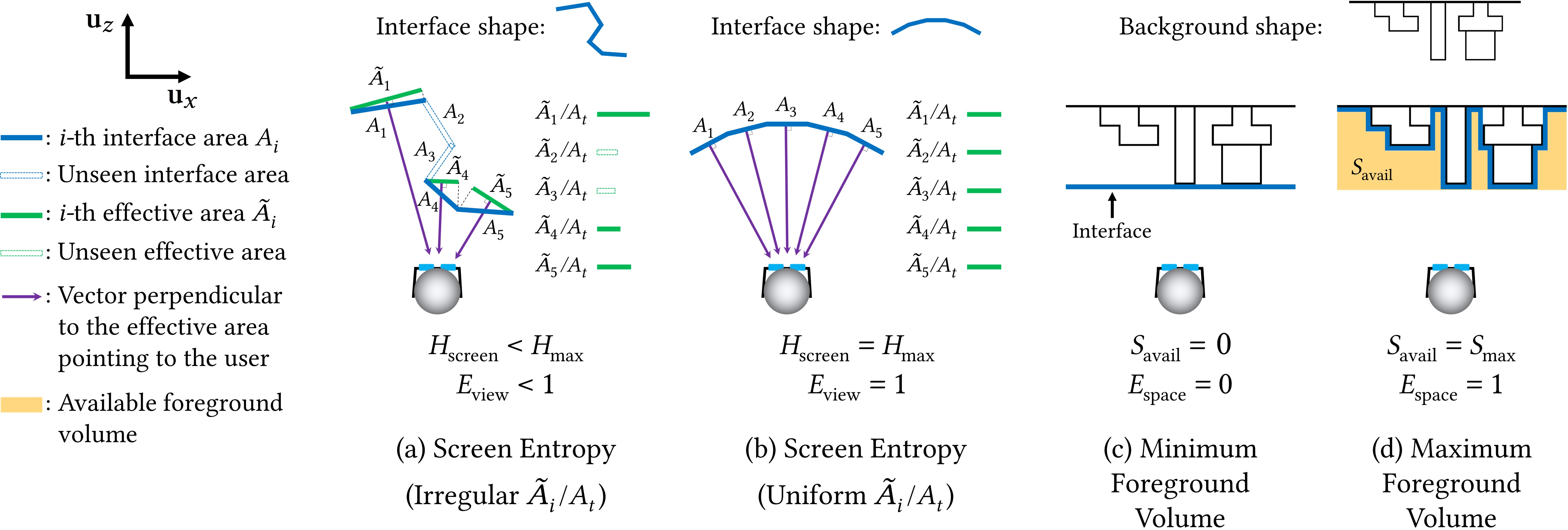}
    \caption{Examples of calculating Screen Entropy ($H_\text{screen}$), View Efficiency ($E_\text{view}$), Available Foreground Volume ($S_\text{avail}$), and Space Efficiency ($E_\text{space}$). (a) The parts of the interface differ in size and orientation, resulting in Screen Entropy below the maximum. (b) The parts of the interface have uniform size and orientation toward the user, resulting in maximum Screen Entropy. (c) The interface overlays the background, leaving empty regions hidden behind it. (d) The interface is tightly fitted to the background, achieving maximum Available Foreground Volume.}
    \Description{This figure shows examples that help explain four measurements used in the paper: how to quantify view quality using screen entropy, which increases when the interface is even and smooth; how view efficiency is defined; how to quantify the available empty space in front of the interface using the sum of all volumes in front of it; and how space efficiency is defined. In part (a), different regions of the interface appear at various sizes and angles. Because these regions are uneven, the screen quality is poor, so view efficiency is low. In part (b), all regions of the interface appear with smooth curvature, and each interface element is oriented directly toward the user. This results in the highest view quality, so screen entropy is maximized and view efficiency equals 1. In part (c), the flat interface is positioned in a way that hides background objects, leaving unused space that the user cannot see. In part (d), the interface fits closely to the background surface, revealing the largest possible amount of available space in front of it. Here, the foreground volume is maximized, and space efficiency equals 1.}
    \label{fig:efficiency_principle}
\end{figure*}

\subsubsection{Screen Entropy and View Efficiency for Uniform Interface Representation}
To address CDR1-(\ref{item:iv2}),
we adapt the concept of viewpoint entropy \cite{vazquez2001viewpoint} to define Screen Entropy $H_{\text{screen}}$. We do this by removing the distance component from the viewpoint entropy and restrict the domain from the entire scene to the screen. $H_{\text{screen}}$ is then normalized by its maximum value $H_\text{max}$ to obtain View Efficiency, $E_\text{view}$. Viewpoint entropy was originally proposed to identify the viewpoint that maximizes information by ensuring that scene faces are uniformly visible \cite{vazquez2001viewpoint, vazquez2003automatic}. In our case, maximum information is obtained when interface mesh elements are oriented
toward the user's viewpoint and have uniform size. Based on this property, we define $H_\text{screen}$ and $E_\text{view}$ as follows:
\begin{equation}
H_\text{screen} = -\sum_{i=1}^{N_f}\frac{\tilde{A}_i}{A_t} \log_2 \frac{\tilde{A}_i}{A_t}\quad E_\text{view} = \frac{H_\text{screen}}{H_{\max}}
\label{eq:screen_entropy}
\end{equation}
where $N_f$ is the number of mesh faces, $\tilde{A}_i$ is the effective area of the $i$‑th mesh face area $A_i$ projected toward the user (Fig.~\ref{fig:efficiency_principle}(a)), and $A_t$ is the sum of all $A_i$. $\tilde{A}_i$ and $A_i$ are considered as 0 if they are not visible to the user. The maximum value of $H_\text{screen}$ is $H_\text{max}=\log_2 N_f$, according to information theory~\cite{shannon1948mathematical}.
As orientation and size uniformity among faces increase, $E_\text{view}$ approaches 1 (Fig.~\ref{fig:efficiency_principle}(b)), providing a quantitative measure of interface uniformity. We consistently applied these values according to the grid resolution established at interface initialization.

\subsubsection{Available Foreground Volume and Space Efficiency for Immediate Environment Understanding}

To address CDR2-(\ref{item:eu1}) and CDR2-(\ref{item:eu2}),
we define the available foreground volume $S_\text{avail}$. 
This is computed from depth differences along the placement line between a planar interface (Fig.~\ref{fig:efficiency_principle}(c)) and a draped interface. 
The maximum value $S_\text{max}$ is obtained from the initial interface before smoothing (Fig.~\ref{fig:efficiency_principle}(d)). 
Based on these values, $E_\text{space}$ is defined as shown in Eq.~\ref{eq:riemann_sum}.

\begin{equation}
S_\text{avail} = \sum_{i=0}^{N_x-1} \sum_{j=0}^{N_y-1} \left(D_{i,j}^\text{n}-D_\text{min}\right), \quad E_\text{space} = \frac{S_\text{avail}}{S_{\max}}
\label{eq:riemann_sum}
\end{equation}
where $N_x$ and $N_y$ are index counts along $x$ and $y$. As the available foreground space increases, $E_\text{space}$ approaches 1, providing a quantitative measure of space efficiency.

\subsubsection{Convergence Condition: Balanced Point and Full Visibility}

To balance uniform surface representation and available foreground volume while considering full visibility of interface face elements, we define the convergence condition using two criteria:
\begin{enumerate}
    \item $E_\text{view} = E_\text{space}$ \hspace{1em} (the iteration number when this is first satisfied is defined as $\text{n}=\text{C}_\textit{B.P.}$)
    \item $\theta_i < 90$° for all $i = 1, 2, \ldots, N_f$ \hspace{1em} (the iteration number when this is first satisfied is defined as $\text{n}=\text{C}_\textit{F.V.}$)
\end{enumerate}
where $\theta_i$ is the angle between the normal of the $i$-th mesh face and the line from the viewpoint to the mesh face center.
The first equality defines the \textit{Balanced Point} (\textit{B.P.}), where the efficiency curves intersect. The second inequality ensures \textit{Full Visibility} (\textit{F.V.}, see CDR1-(\ref{item:iv1})) by eliminating cases where invisible regions remain, as may occur when parts of the interfaces are too close or partially folded. For the second criterion, we assume two viewpoints approximately 3 cm to the left and right of the original viewpoint, based on the interpupillary distance (IPD) \cite{dodgson2004variation}, and compute $\theta_i$ for both eyes. Therefore, the final convergence iteration is given by
\begin{equation}
\text{C} = \max(\text{C}_\textit{B.P.}, \text{C}_\textit{F.V.})
\label{eq:cmax}
\end{equation}
This condition guarantees that every convergence condition achieves full visibility at the viewpoint used for CAMEO interface placement, including the case when $\text{C}_\textit{B.P.}$ is chosen. This is because when $\text{C}_\textit{B.P.} > \text{C}_\textit{F.V.}$, $\text{C}_\textit{B.P.}$ already achieves full visibility. If $\text{C}_\textit{F.V.}>\text{C}_\textit{B.P.}$, the interface instead converges to a state where $E_\text{view} > E_\text{space}$.

\begin{figure*}[h]
    \centering
    \includegraphics[width=\textwidth]{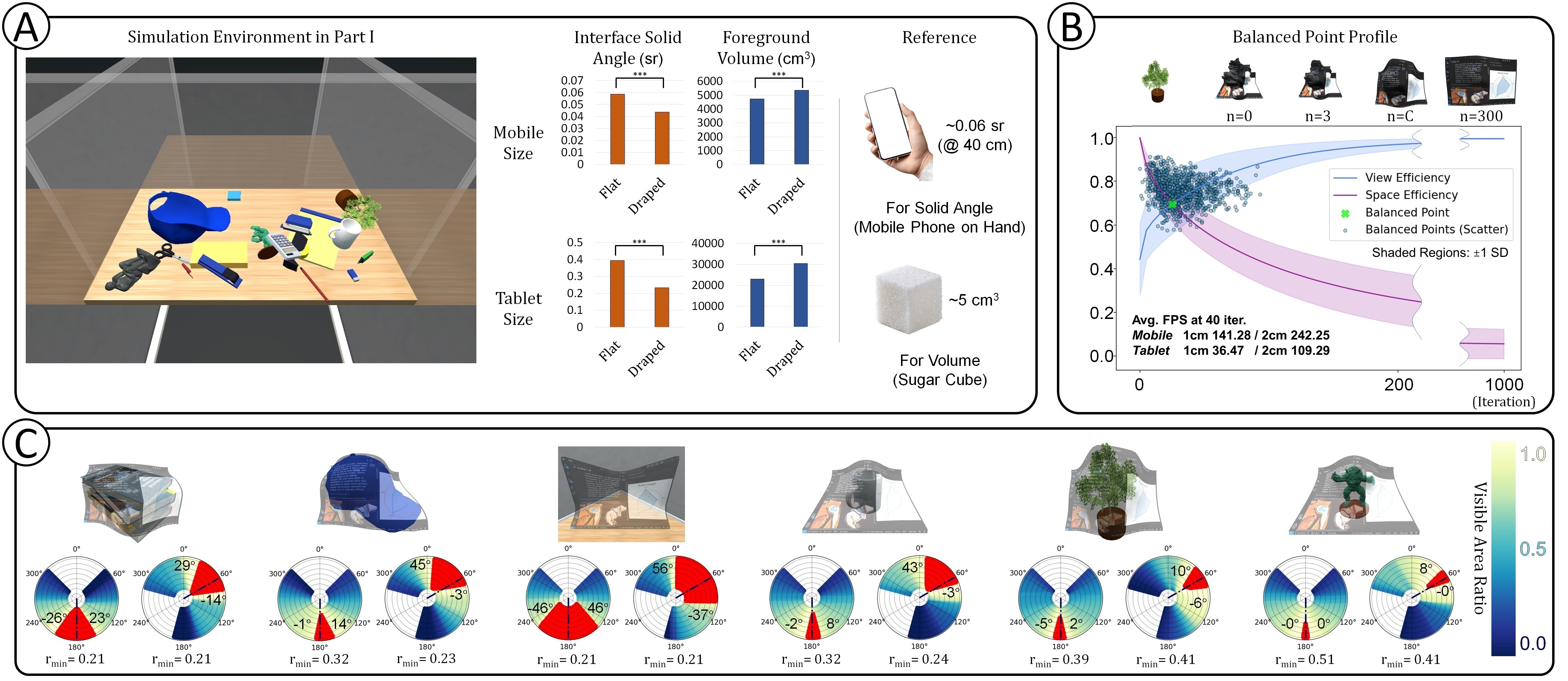}
    \caption{\CHG{\textbf{A (Part I)}: Example scene of the simulation environment and differences in interface occlusion and foreground volume across the interface shapes. Asterisks (***) indicate $p < .001$.
    \textbf{B (Part II)}: Example of two efficiency curves. The curves were drawn using 6 selected objects. Blue scatter points indicate the Balanced Points from the simulation runs for the tablet size in Part I. The example images above the graph show the degrees of draping formed on the Plant object across four iterations. FPS is reported for each interface size and vertex density.
    \textbf{C (Part III)}: Left and right panels show front and side views, with the placement line in blue dashes. Heatmaps mark fully visible (100\%) regions with added red shading. Angles indicate the maximum full-visibility range at $r = 0.5$\,m, and the minimum radial distance for full visibility is $\mathrm{r}_{\mathrm{min}}$ (m). Please zoom in for clarity. See Appendix C for further details.}}
    \Description{Part A (Simulation Part One) This section shows an example of the simulation environment and demonstrates how occlusion and foreground volume change for mobile-sized and tablet-sized interfaces in both flat and draped conditions. The results show that, for both sizes, draping significantly reduces occlusion and significantly increases foreground volume. In addition, larger interface sizes consistently secured more foreground volume and achieved greater occlusion reduction in both interface conditions. Part B (Simulation Part Two) This section shows curves representing how the interface changes over smoothing iterations. The balance points for the tablet-sized simulations from Part One are scattered around iteration 40 and an efficiency of approximately 0.75. Above the graph, four example images of the Plant object with the draped interface illustrate how the interface conforms more tightly or loosely to the object as the number of smoothing iterations increases. Near 40 iterations, FPS performance results are shown for each interface size and vertex grid density: on mobile, 141.28 FPS for the 1 cm grid and 242.25 FPS for the 2 cm grid; on tablet, 36.47 FPS for the 1 cm grid and 109.29 FPS for the 2 cm grid. Part C (Simulation Part Three) This section provides radial heatmaps that show how much of the interface area becomes visible from different user viewpoints, visualized from both front-view and side-view perspectives. Six example objects are shown. The results indicate that more of the interface becomes visible as the distance along the placement line increases. Viewpoints from which the entire interface is visible are distributed in a fan-like pattern near the placement line. The figure also reports the minimum distance at which the entire interface becomes fully visible. Please refer to Appendix C for further details.   
    }
    \label{fig:simulation_result}
\end{figure*}

\CHG{We adopt \text{C} as our baseline in subsequent experiments, as it offers a reproducible setting between the two conflicting constraints and provides an interpretable reference. However, in real applications, users can still adjust the smoothing level to accommodate \CHG{environmental} understanding and interface legibility, depending on task demands and individual preferences.}

\section{Simulation Studies: Immediate Environment Understanding and Interface Shape Changes}
\label{sec:study_i}

\CHG{We conducted a three-part simulation study to evaluate how draping affects occlusion, available foreground space, deformation behavior, and interface visibility across viewpoints to provide quantitative grounding in both geometric and performance aspects prior to running user evaluations. All simulations were implemented in Unity 2022.3.9f1 on a desktop workstation equipped with an NVIDIA GeForce RTX 3060 GPU, an Intel Core i5-12600K CPU, 16 GB of RAM, and Microsoft Windows 11 Enterprise 64-bit. For \textbf{Part I}, our simulation study focuses on two key workspace properties: background occlusion, which can impair performance and increase risks~\cite{kiral2024evaluating,boothroyd2010product,kiral2023assessing}, and foreground volume, which indicates accessible space and supports efficient interaction~\cite{sivak1990integration}. We quantified occlusion (using solid angle~\cite{BIPM2019}) and the available foreground volume across the Flat and Draped conditions (iteration $\text{n}=\text{C}$; Eq.~\ref{eq:cmax}). For this analysis, we used tablet-(30 cm~$\times$~20 cm) and mobile-(15 cm~$\times$~7 cm) sized interfaces to reflect practical form factors commonly used in everyday contexts. These interfaces are smaller than typical monitors, making them plausible candidates for MR interfaces with minimal obstruction of the environment~\cite{davari2020occlusion}. All conditions were examined in both portrait and landscape orientations. We also examined the extent to which the convergence condition was satisfied by either $\text{C}_\textit{B.P.}$ or $\text{C}_\textit{F.V.}$ (Eq.~\ref{eq:cmax}). Using 8,000 simulated trials across 200 randomized desk environments populated with 17 randomly placed office items (e.g., scissors, stapler, notepad, etc.) (Fig.~\ref{fig:simulation_result}(A) left) and randomly varied viewpoints reflecting posture variability~\cite{brink2014spinal}, the results showed that Draped interfaces occluded less of the visual field and created more available foreground volume (Fig.~\ref{fig:simulation_result}(A) middle). 
Convergence analysis further indicated that in nearly all cases the Balanced Point condition ($\text{C}_\textit{B.P.} > \text{C}_\textit{F.V.}$) was the dominant criterion (98.45\% of cases were $\text{C}=\text{C}_\textit{B.P.}$), confirming that the Balanced Point operates reliably for most Draping configurations.
\textbf{Part II} examined how increasing the iteration count n in iterative smoothing affects the interface shape focusing on a tablet-sized interface, which is capable of supporting diverse usage scenarios, and additionally measured system performance around the balanced point ($\text{n}=40$) for reference. To observe shape transformations, we used six objects with diverse geometries (Books, Cap, Corner, Mug, Plant, and Statue), chosen for their concave--convex structures and depth variations. For performance measurement, we recorded frame time for 1 minute across tablet and phone interface sizes, testing both 1 cm and 2 cm vertex grid densities. The results show that increasing the number of iterations produced flatter interfaces, which increased occlusion and reduced foreground volume (Fig.~\ref{fig:simulation_result}(B) top; see Appendix C.3, C.4 for details). In terms of performance, the system supported interactive frame rates around the balanced point. The results also revealed that larger interfaces with denser vertex grids required greater computational resources, indicating areas for future work to enhance interface fidelity and support always-responsive level of operation. (see FPS data in Fig.~\ref{fig:simulation_result}(B)). The middle of Fig.~\ref{fig:simulation_result}(B) shows the trade-off between View Efficiency and Space Efficiency across iterations, with a balanced point emerging at their intersection ($\text{n} = C_{\textit{B.P.}}$).
\textbf{Part III} evaluated whether Draped interfaces at the Balanced Point remain visible under variation in viewpoint. For each of the six object shapes, more than 33k viewpoints per interface were sampled. The results showed that once full visibility is achieved along the placement line, it persists across extended regions as the viewer moves farther away (Fig.~\ref{fig:simulation_result}(C)). Moreover, the continuous change in visibility-area ratio across viewpoints reflects how the inherent geometry of the Draped interface influences its visibility characteristics.}

\section{User Study I: Environmental Interaction in Mixed Reality Workspace}\label{sec:study_ii}

\CHG{We investigate whether \CHG{CAMEO} can reduce the interference on motor movement, associated with planar MR windows, when reaching for nearby objects. Conventional planar interfaces can block sight lines, which may alter motor performance~\cite{sivak1990integration}. To assess how interface geometry influences visually guided action, we use a reaching task with targets placed behind or around the UI. } As a baseline, we compare \CHG{CAMEO} against flat, cockpit-style MR interfaces~\cite{ens2014personal, li2024situationadapt} to assess the extent of hand movement required when reaching for surrounding objects. We presume that larger movements indicate greater interference. Our research questions are as follows:

\begin{enumerate}
    \item[\textbf{RQ1:}] How do draped versus flat interface layouts affect hand movement when interacting with surrounding objects?
    \item[\textbf{RQ2:}] How do task load, usability, and \CHG{environmental} understanding differ between draped and flat layouts, with and without transparent windows?
\end{enumerate}

\subsection{Apparatus and Experimental Setup}

\begin{figure}[htbp]
    \centering
    \includegraphics[width=\columnwidth]{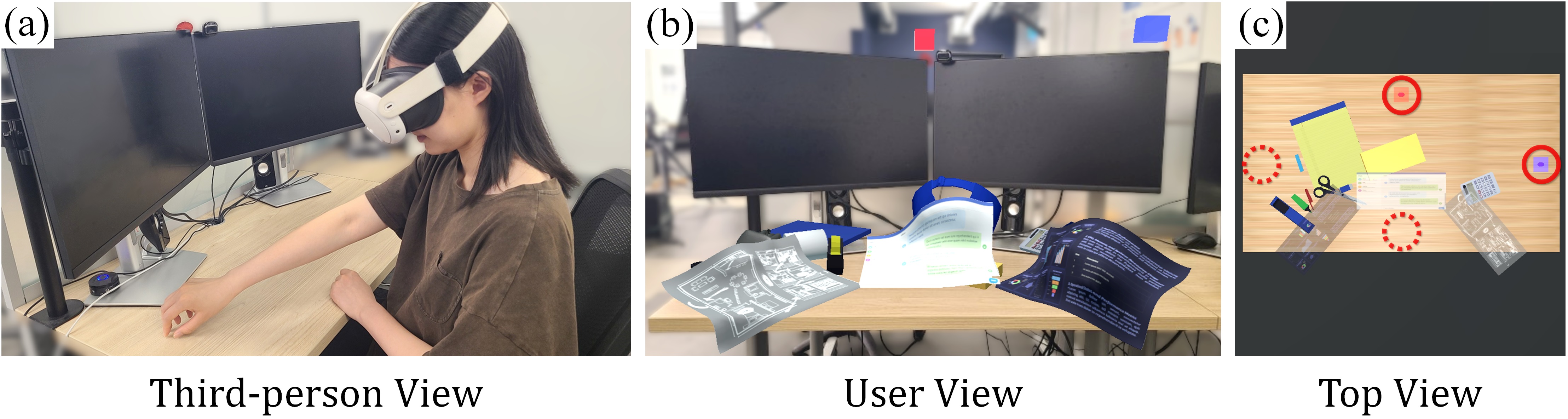}
    \caption{\CHG{The environment for interacting with nearby targets. (a) The real setup where the user is selecting targets. (b) An MR user view in the Opaque Draped (\textit{O.D.}) with the \textit{Heavy} condition. Mid-air cubes mark target locations to support locating the targets which could be hidden behind the interfaces. (c) Two activated targets (solid circles) and the inactivated region (dashed circle) are shown in the Transparent Flat (\textit{T.F.}) with \textit{Light} condition.}}
    \Description{This figure shows an example of the environment used for the interaction task. In part (a), the image shows the real physical setup that participants interact with: an empty office desk with two desktop monitors, along with an example participant wearing an HMD. In part (b), the image shows what a user sees through the mixed-reality headset when using the opaque draped interface under the heavy environmental setting. Floating cubes in mid-air guide the locations that the user must reach during the task. In part (c), the top-view image shows two active target locations on the desk. This example view is displayed using the transparent flat interface under the light setting.}
    \label{fig:study1_env}
\end{figure}

\CHG{We conducted the experiment in an MR environment using the video see-through mode of the Meta Quest 3 on a real desk (100 cm × 60 cm, height 72 cm). Two monitors were placed on the desk (Fig.~\ref{fig:study1_env}(a)), and virtual objects and three MR interfaces were displayed (Fig.~\ref{fig:study1_env}(b), (c)). The six virtual objects were used to precisely control geometry and occlusion conditions, consistent with Simulation Part I (Section~\ref{sec:study_i}). We created six environments by varying the number of office items, consisting of \textit{Heavy} (Fig.~\ref{fig:study1_env}(b); all 17 objects) and \textit{Light} (Fig.~\ref{fig:study1_env}(c); only 10 relatively small items such as a pen or a notepad sets with three variations each). See Appendix C.1 and D.1 for details.}

Targets were represented as small 3D \CHG{bead-shaped objects}. The workspace included three tablet-sized windows (30 cm × 20 cm), chosen to reflect typical MR window usage. \CHG{This tablet-scale dimension was selected because it can reflect practical window sizes that can support versatile application in desktop computing scenarios.}
The central interface was positioned along the line from the user’s viewpoint to the desk, and the other two were rotated relative to the center. Interface content types were randomly assigned among Dashboard, Chat screen, and Storyboard to account for variations in brightness. For transparency, the $\alpha$ value was set to 0.5 following recommendations from prior work~\cite{hussain2024effect}.

\subsection{Participants}

Sixteen participants (female: 8, male: 8) were recruited via online platforms and through posters at our institution and in local communities. Their average age was 26.2 years (SD = 5.14, range = 19–39). Participants provided informed consent and received about 20 USD. \CHG{The study was approved by our institution's Research Ethics Board.} Participants showed varied frequencies of VR HMD use: 2 had never used an HMD, 6 used it rarely, 3 used it several times a month, 3 several times a week, and 2 daily. All except one were right handed.

\subsection{Movement Task}

Participants selected background bead-shaped targets placed along the horizontal and vertical midlines, each 5 cm from the desk edges. In each trial, two targets appeared (Fig.~\ref{fig:study1_env}(c)). A red cube marked the start target, and a blue cube marked the end target.
To cover all paths between the four targets in both directions, a random Euler circuit~\cite{bondy1976graph} was generated from a complete symmetric digraph. This defined a task path of 12 movements per hand, yielding 24 trajectories in total.
A grasp was registered when a fingertip approached within 3 cm of a target, accompanied by an audio cue. The starting hand alternated by condition, and participants switched hands after 12 movements. Targets were positioned 2 cm above the desk to avoid unintended collisions.

\subsection{Study Design}

We used a within-subjects design with \textbf{Interface Shape} and \textbf{Transparency} as factors, resulting in four conditions: \textit{Opaque Flat (\textbf{O.F.})}, \textit{Opaque Draped (\textbf{O.D.})}, \textit{Transparent Flat (\textbf{T.F.})}, and \textit{Transparent Draped (\textbf{T.D.})}. Conditions were grouped into \textit{Transparent} and \textit{Opaque} blocks. Block order and shape order were fully counterbalanced across eight sequences, with two participants assigned to each sequence while balancing gender distribution. A without-interface condition \textit{(\textbf{W/O})} was also included at varied points in an evenly distributed manner to reduce order effects. Within each condition, participants experienced both \textit{Heavy} and \textit{Light} environments in randomized order.

\subsubsection{Procedure}

After a task overview, participants wore a Meta Quest 3 HMD, sat at the desk, and completed practice trials. They performed pinch gestures to grasp targets and followed the start end cues indicated by colored cubes. To ensure accurate measurement from start to end, participants were required to identify the designated start and end points first through the cubes.  
Participants were verbally instructed to perform the task as conveniently and directly as possible, and to remain consistent at all times. Regarding virtual object collisions, they were told that the interface could be touched or penetrated, whereas objects themselves should not be passed through. After each condition, participants rested and completed questionnaires except in the without-interface condition.

\subsubsection{Measures}

Objective measures included hand movement derived from the palm position tracked by the vision based tracking system of the Meta Quest 3 (v78). To reduce the complexity of individual postures and limit the degrees of freedom from finger movements during measurement, palm positions were used as movement indicators, defined as the center of the metacarpals. In addition, the duration of hand penetration into objects was measured to confirm whether participants avoided object collisions. We also measured the duration of hand penetration into objects. Trials were marked as outliers if participants switched hands during a movement or if tracking failed. To capture all of the participant’s unique hand movements, we only classified a trial as an outlier when the initiating hand deviated by more than three times the interquartile range (IQR). We resumed measurement if participants re-touched the start target. We excluded trials from the analysis if hand-tracking failure resulted in missing data. Subjective measures included the NASA-TLX~\cite{hart1988development} for Task Load and the pragmatic quality subscale of the UEQ-S~\cite{hinderks2017design} for Usability. To assess participants’ environmental understanding and their preferences, they rated the two customized questions: \textit{``In this condition, how well could you understand the environment during the task?''} and \textit{``How much do you prefer this condition?''}, using a 7-point Likert scale. Additionally, to capture their general impressions, we collected their comments.

\subsection{User Study I: Results and Discussion}

A total of 4.59\% of trajectory data was excluded due to outliers or missing data. Exclusion rates by condition were \textit{W/O} = 3.95\%, \textit{T.D.} = 3.98\%, \textit{O.D.} = 4.41\%, \textit{T.F.} = 4.82\%, and \textit{O.F.} = 5.78\%. Average task completion time was 8.1 minutes per condition (SD = 0.8373). Average collision time was 3.04\% (SD = 1.7348).

\begin{figure*}[htbp]
    \centering
    \includegraphics[width=\textwidth]{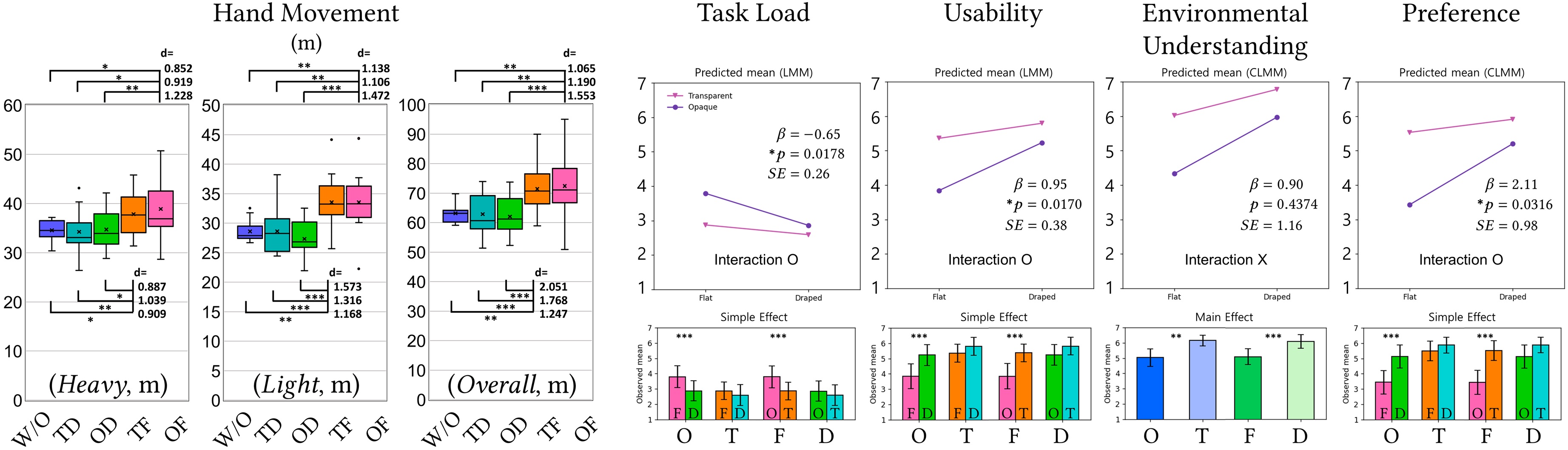}
    \caption{\CHG{Results of User Study I. For Hand Movement, the box plot includes Bonferroni-corrected significance and Cohen's d; the $\times$ indicates the mean and the horizontal line indicates the median. For subjective measures, interaction plots display interaction estimates ($\beta$) as interaction effect-size indicators, $p$-values, and standard errors ($SE$). Auxiliary bar charts based on observed means are provided as visual aids with 95\% CIs. With significant interactions, they show simple effects; otherwise main effects. Holm-corrected $p$-values were applied to all multiple comparisons. Asterisks indicate significant interactions, main effects, or simple effects based on Wald t/z tests. * $p$ < .05, ** $p$ < .01, *** $p$ < .001. See Appendix D.2 for more detailed data.}
}
    \Description{This figure presents the results of User Study One. For the hand-movement measurements, the box plot shows which differences are statistically significant after applying the Bonferroni correction. It also reports effect sizes using Cohen’s d. For the subjective-response measures, the interaction plots show how two factors influence each other. These plots display the estimated size of the interaction effect, the probability value indicating statistical significance, and the standard error. Additional bar charts, based on the observed average values, are included as visual guides and show ninety-five percent confidence intervals. When there is a significant interaction between factors, the charts show the simple effects; when there is no significant interaction, they show the main effects instead. All multiple comparisons use probability values corrected with the Holm method. Asterisks mark statistically significant findings, whether they come from interactions, main effects, or simple effects, based on Wald tests. One asterisk indicates significance below point zero five, two asterisks indicate significance below point zero one, and three asterisks indicate significance below point zero zero one. Additional details can be found in Appendix D point two.}
    \label{fig:study2_results}
\end{figure*}

For hand movement (Fig. \ref{fig:study2_results} left), a one-way repeated-measures ANOVA (RM-ANOVA) showed significant differences across conditions, reported as ($F(4,60)$, $p$, $\eta^2_p$), for \textit{Heavy} (9.43, $<0.001$, $.386$), \textit{Light} (19.79, $<0.001$, $.569$), and \textit{Overall} (21.97, $<0.001$, $.594$) sets, so we subsequently conducted pairwise analyses. Pairwise comparisons indicated that \textit{Flat} interfaces resulted in significantly longer hand movement than \textit{Draped} or \textit{W/O} settings. This suggests that \textit{Flat} layouts led to greater detours. Effect sizes for significant cases are presented in Fig.~\ref{fig:study2_results} left.
\begin{table*}[h!]
\centering
\caption{Pairwise PERMANOVA results for trajectory differences between conditions in the interaction task. Each trajectory difference between conditions assessed based on Hausdorff distances.
Values represent pairwise pseudo-F statistics, with numbers in parentheses listed according to pathway numbers (Fig.~\ref{fig:trajectory}). 
Bonferroni correction was applied; * indicates $0.01 < p_c < 0.05$, otherwise $p_c = 0.01$ (based on 999 permutations), \CHG{indicating that all presented paths in the table were significantly different.}}
\resizebox{\textwidth}{!}{
\begin{tabular}{lccc}
\toprule
Condition & Without Interface (\textit{W/O}) & Transparent Draped (\textit{T.D.}) & Opaque Draped (\textit{O.D.}) \\ \hline
 Transparent Flat (\textit{T.F.}) & 
(10.62, 8.93, 19.59, 10.50, 21.20, 16.88) & 
(9.03, 6.60, 11.64, 6.01$^*$, 13.46, 8.84) & 
(14.20, 10.50, 15.58, 12.84, 18.11, 14.03) \\
Opaque Flat (\textit{O.F.}) & 
(15.23, 7.82, 30.54, 10.99, 31.00, 20.09) & 
(13.08, 6.32, 20.10, 10.46$^*$, 20.19, 10.51) & 
(21.14, 10.89, 25.38, 13.73, 27.09, 18.75) \\ \bottomrule
\end{tabular}
}
\label{tab:permanova}
\end{table*}

\begin{figure}[htbp]
    \centering
    \includegraphics[width=\columnwidth]{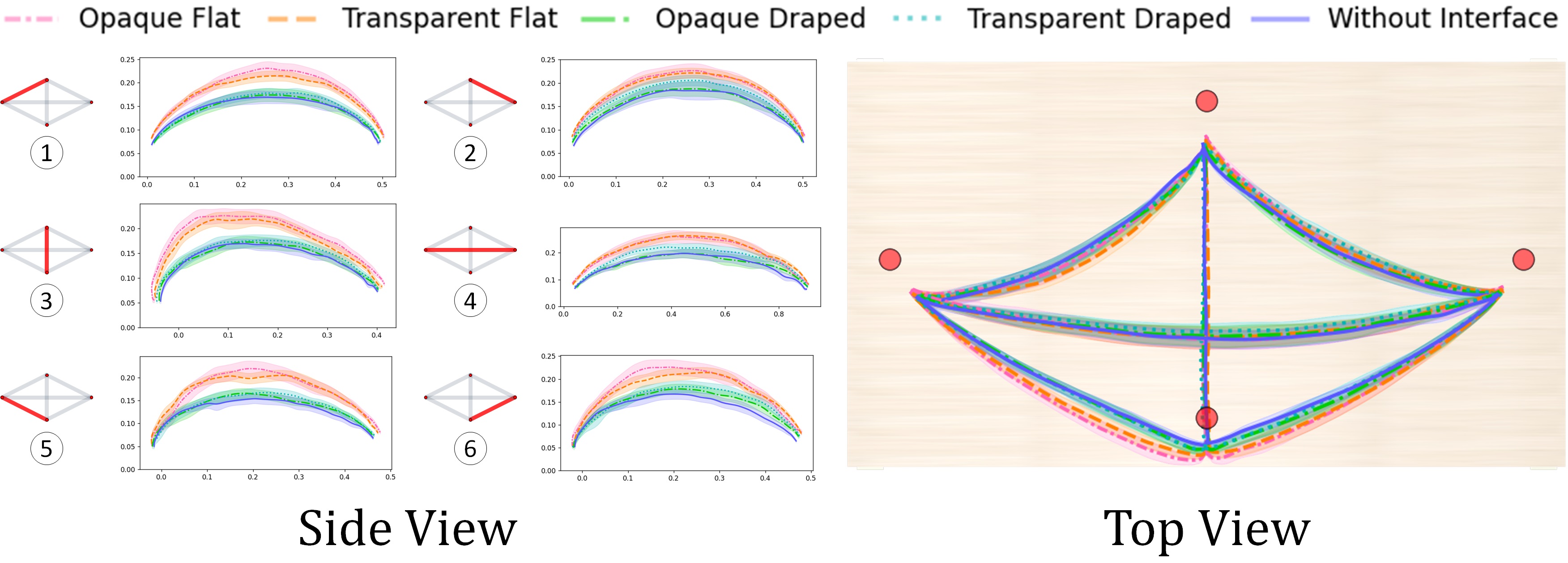}
    \caption{Hand movement trajectory. For each path, the mean trajectory was projected onto the vertical plane (Side View) and horizontal plane (Top View). The scale ratio is 1:1 for both axes. In the Top View, the desk target locations are indicated by red circles. The shaded area represents the 95\% confidence interval of the projected trajectories. Based on these shaded area, the detour effect is more clearly observed in the Side View. In the Side View, the paths between targets and their numbers are displayed on the left side of each graph.}
    \Description{This figure shows the arc-shaped movement paths of the hand during the task. The shaded area represents the 95 percent confidence interval of the projected trajectories, allowing the variation in participants’ movements to be estimated. For each movement path, the average trajectory was projected onto two views: a vertical plane, shown as the side view, and a horizontal plane, shown as the top view. In the top view, the target locations on the desk are shown as red circles. The side view makes the detour effect easier to observe.}
    \label{fig:trajectory}
\end{figure}

To further examine movement patterns, we analyzed 3D trajectories. Trajectories were interpolated to equal lengths, and reversed when necessary for comparison. Fig. \ref{fig:trajectory} shows that \textit{Flat} conditions produced larger vertical deviations than \textit{Draped} conditions. We quantified trajectory differences using the Hausdorff distance~\cite{alt2009computational}, a commonly used measure in trajectory analysis~\cite{bian2019trajectory}, and tested significance with Permutational Multivariate ANOVA (PERMANOVA) \cite{bian2019trajectory}. Differences across conditions were significant ($p$ = 0.001, the number of permutations = 999; \CHG{global} pseudo-F values: Path 1: 8.749, Path 2: 5.830, Path 3: 13.769, Path 4: 7.558, Path 5: 14.649, Path 6: 10.034; see path numbers in Fig.~\ref{fig:trajectory}), \CHG{so we proceeded with pairwise comparisons, and the results show that the \textit{Flat} conditions were significantly different from all other conditions across all paths (Table~\ref{tab:permanova}), suggesting that the \textit{Flat} setting may have contributed to detour trajectories.}

\CHG{For the subjective measures (Fig.~\ref{fig:study2_results} right), Task Load and Usability were analyzed with linear mixed models (LMMs) since averaging NASA-TLX/UEQ-S items yields continuous outcomes with no well-defined ordinal categories or cut-points. This approach has also been adopted in prior work~\cite{kumar2025mixed, harris2020development}. By contrast, \CHG{Environmental} Understanding and Preference, which were ordinal outcomes, were analyzed using cumulative link mixed models (CLMMs) to appropriately preserve the ordinal scale. For Task Load, Usability, and Preference, the interactions between factors were significant, meaning that Interface Shape and Transparency influenced each other, so we analyzed them using simple effects. For \CHG{Environmental} Understanding, the interaction was nonsignificant, so we analyzed it using main effects.
For both Task Load and Usability, \textit{Draped} was less demanding and more usable than \textit{Flat} in the \textit{Opaque} condition, while within the \textit{Flat} interface, the \textit{Transparent} condition was associated with lower workload and higher usability than the \textit{Opaque} condition.
For \CHG{Environmental} Understanding, both \textit{Draped} and \textit{Transparent} conditions independently helped users understand the background more clearly.
For Preference, \textit{Draped} maintained high preference even without transparency, whereas in the \textit{Flat} interface, preference dropped sharply when transparency was absent. Taken together, transparent displays and our draped method offer complementary benefits that can be combined to expand user options.}

Participants’ subjective comments provide further insight into how Interface Shape and Transparency influence task performance. Participants consistently reported discomfort with flat interfaces, describing them as physically demanding, intrusive, and visually obstructive. \textit{Opaque Flat} conditions in particular were frequently labeled “annoying (P02),” “overwhelming (P08),” or task-blocking (P04, P09, P10, P12, P16). In contrast, \textit{Draped} interfaces were generally preferred for reducing occlusion (P02, P08, P11, P15) and supporting task performance (P03, P05, P09, P10, P12, P15). \textit{Transparent Draped} displays were often seen as the most effective combination, as they minimized blocking while preserving background visibility: “Even transparent drape is much better because it blocks less” (P11). However, preferences were not unanimous; some users experienced dizziness (P07) or misperceived draped displays as physical obstacles, leading to avoidance behaviors (P13). Transparency itself was widely valued for improving environmental understanding (P03, P04, P09, P10, P13, P15, P16), though issues such as “double vision” (P10) and “glass-like” avoidance (P07) were reported. Overall, \textit{Draped} conditions alleviate many limitations of \textit{Flat} displays, particularly \textit{Opaque Flat}, while participants’ perceptual challenges reveal new insights into how users engage with semi-trasparent elements.

\subsection{User Study I: Limitations}

This study aimed to demonstrate the advantages of \CHG{CAMEO} over planar interfaces in terms of space efficiency and environmental adaptability. However, our baseline (\textit{Flat} cockpit setting) may not fully reflect optimized planar layout strategies, as skilled users or optimization techniques~\cite{li2024situationadapt, cheng2023interactionadapt} can reduce unused space. Also, in completely flat environments with no space behind the interface, the space-efficiency benefits of our method may diminish. Even so, planar interfaces remain inherently limited in placement flexibility and spatial efficiency, often relying on a single optimal placement~\cite{li2024situationadapt, cheng2023interactionadapt}. Therefore, we recommend interpreting this as meaning that our method is most beneficial in complex environments unsuited to flat interfaces.

\CHG{Also, although we varied the environment, our study remains limited by the fact that we did not examine all possible interface sizes. This choice was to study an interface that can be used broadly while keeping the experimental complexity manageable. Exploring larger or differently shaped interfaces could reveal a variety of occlusion effects and movement behaviors.}

During the task, participants received only the essential instructions and their hand movements were not strictly constrained, a choice intended to preserve ecological validity. Furthermore, our MR environment used virtual objects for precise and reproducible occlusion control, which may not fully capture the semantic qualities of real physical objects. Motor behavior is influenced by perceptual~\cite{regan2000visually}, emotional~\cite{braine2023emotion}, and biomechanical factors~\cite{lommertzen2008collision}. In future work, investigating these influences may provide additional insights into task performance. 

\CHG{For implementation, our method assumes the availability of accurate meshes for physical objects. This can be achieved through commercially supported HMD-based scanning or customized mesh reconstruction pipelines~\cite{farshian2023deep}. Future work should investigate how such an integrated system performs with our method, with particular attention to the practical constraints it may impose on users’ interactions.}

In our study, the hedonic quality~\cite{hassenzahl2001effect} was not addressed, despite its link to usability~\cite{tractinsky2000beautiful}. User feedback indicated that draped interfaces were sometimes perceived as object-like, “weird,” or distracting. These perceptions offer additional insights into how interface form shapes experience and imply the need for further investigation. Finally, our evaluation covered two static single user environments (\textit{Heavy} and \textit{Light} environments). Broader contexts including dynamic or multi user scenarios could be explored in future work.

\section{User Study II: Interface Legibility and Shape Awareness}\label{sec:study_iii}

In this study we investigate whether draping supports both content legibility and users’ ability to recognize and interact with background objects. 
\CHG{Specifically, there is an inherent tension between legibility and the ability to infer hidden background shapes. A planar surface could offer optimal legibility~\cite{wei2020reading} but provides no cues about the geometry behind it. Conversely, a fully draped interface could support inferences about partially veiled shapes by leveraging natural perceptual abilities~\cite{phillips2020veiled, wong2023seeing}, but this comes at the cost of reduced legibility. To capture this trade-off, we asked participants to perform both a legibility task (Task-\readText{\textbf{Read}}) and a shape-inference task (Task-\matchText{\textbf{Match}}). Our goal is to determine whether environment-adaptive UI geometry can convey contextual information beyond what planar interfaces provide while still maintaining the clarity required for effective interface use.}
To this end, we compare different degrees of draping to assess how they impact reading clarity and shape awareness. From this, we examine the following research questions (RQs):

\begin{enumerate}
    \item[\textbf{RQ1:}] How do participants perform when reading content on Drap\-ed and Flat UI?
    \item[\textbf{RQ2:}] How does draping affect users’ ability to perceive and recognize background objects?
\end{enumerate}

\subsection{Apparatus and Experimental Setup}

\begin{figure}[htbp]
    \centering
    \includegraphics[width=\columnwidth]{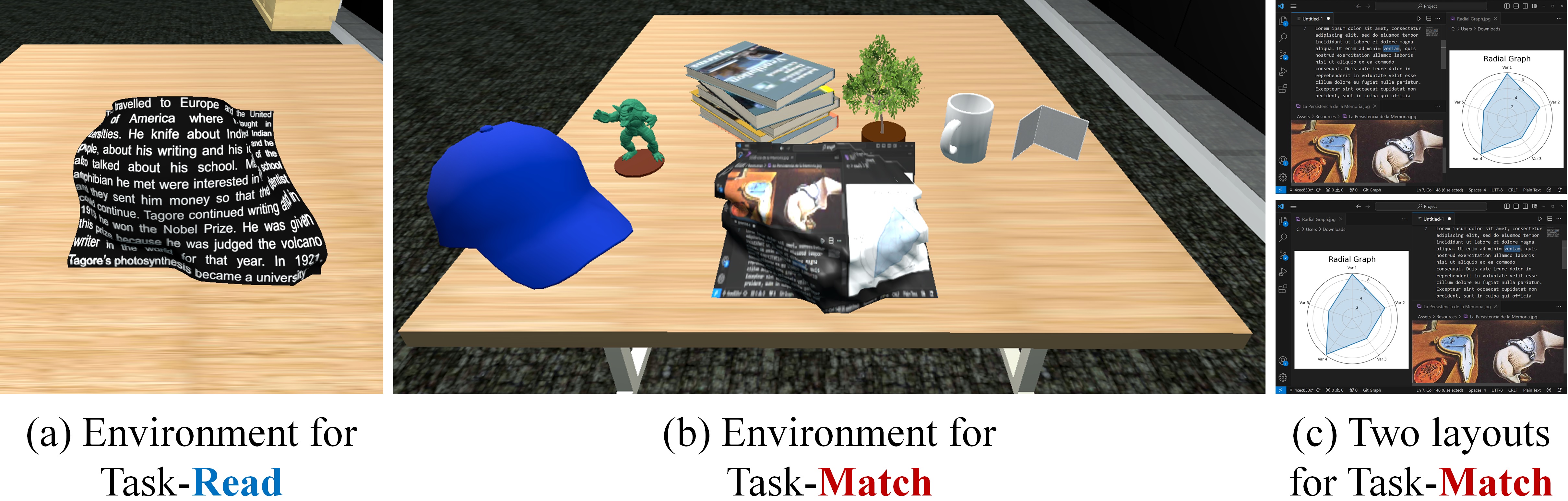}
    \caption{Test environments for each Task-\readText{\textbf{Read}} / Task-\matchText{\textbf{Match}} in User Study II. (a) The empty desk environment setup. This example shows an interface shape draped over the Cap under the \textit{Snug} condition. For the legibility evaluation, participants are asked to find the words in the text that have been intentionally replaced with incorrect ones. (b) 6 objects are placed for a 6-multiple-choice matching task. The corner shape (the rightmost one) is represented by two scaled-down partition walls. The interface shape shown here is created by draping over the Books under the \textit{Tight} condition. For the guessability evaluation, participants identify the object behind the interface. (c) Two interface layouts shown in the matching task. The bright positions are configured differently in each.
    }
    \Description{This figure shows the test environments used for the reading task and the matching task in User Study Two.  In part (a), the scene shows a virtual empty desk used for the reading task with the draped interface. The example demonstrates an interface shape that has been draped over a cap in the condition where the interface fits snugly to the object. In this task, participants evaluate legibility by reading a block of text and locating the words that have been intentionally replaced with incorrect ones.  In part (b), six physical objects are placed on the desk for the matching task, where participants must choose which object matches the shape they see. The object representing a corner shape, located on the far right, is shown using two small partition walls placed together. The interface shape shown here is created by draping over a stack of books in the condition where the interface tightly conforms to the object. In this task, participants judge guessability by identifying which object is hidden behind the interface.  In part (c), two different interface layouts used in the matching task are shown. To add diverse conditions, the layouts differ in the placement of bright and dark regions.}
    \label{fig:test_environments}
\end{figure}

\CHG{To control interruptions that may arise from the surrounding environment during reading tasks ~\cite{chevet2022breaks}, the experiment was carried out in a virtual office environment using the Meta Quest 3 headset and Unity 2022.3.9f1.} The virtual space reproduced a desk of 72 cm height and 100 cm × 120 cm dimensions. 
Participants were positioned centrally in front of the desk and could move their heads freely while seated. Ceiling lights around the user replicated the actual environment, and specular lighting was removed from the interface surface. A custom shader was used to ensure the interface appeared first.
6 objects (Books, Cap, Corner, Mug, Plant, and Statue) were selected to capture diverse shapes, consistent with Section~\ref{sec:study_i}.
\CHG{Using the same 30 cm × 20 cm interface size as in previous studies, we draped it over the six objects to generate 24 interface shapes, with 4 Draped UIs per object created using different smoothing iteration numbers n.}

\begin{itemize}[label=]
    \item \textit{(UI condition)} \readText{$\blacksquare$}: Used for Task-\readText{\textbf{Read}}
    \item \textit{(UI condition)} \matchText{$\blacksquare$}: Used for Task-\matchText{\textbf{Match}}
\end{itemize}

\begin{enumerate}
  \item \textbf{\textit{Tight}} \matchText{$\blacksquare$}: Preserves all surface information ($\text{n}=0$).
  \item \textbf{\textit{Snug}} \readText{$\blacksquare$}\matchText{$\blacksquare$}: Slightly smoothed, but surface information largely preserved ($\text{n}=3$).
  \item \textbf{\textit{Balanced Point (B.P.)}} \readText{$\blacksquare$}\matchText{$\blacksquare$}: A balance between smoothing and surface shape preservation ($\text{n}=\text{C}$; Eq.~\ref{eq:cmax}).
  \item \textbf{\textit{Loose}} \readText{$\blacksquare$}\matchText{$\blacksquare$}: Nearly flat, while still retaining some surface information ($\text{n}=300$).
  \item \textbf{\textit{Flat}} \readText{$\blacksquare$}: No surface information; corresponds to an infinite number of iterations ($\text{n}$~→~$\infty$).
\end{enumerate}

For each object, the number of iterations \text{C} under the \textit{B.P.} condition is as follows: Books 22, Cap 13, Corner 17, Mug 29, Plant 46, and Statue 39. For Task-\readText{\textbf{Read}}, we compared \textit{Snug}, \textit{B.P.}, \textit{Loose}, and \textit{Flat}. For Task-\matchText{\textbf{Match}}, we used \textit{Tight}, \textit{Snug}, \textit{B.P.}, and \textit{Loose}. The draping assumed a desktop viewing posture at a downward angle of $30^\circ$.
Both reading and matching tasks displayed the interface on the desk Fig.~\ref{fig:test_environments}(a, b). In Task-\matchText{\textbf{Match}}, the 6 candidate objects were placed on the desk, and two interface layouts were applied to add diversity (Fig.~\ref{fig:test_environments}(c)).

\subsection{Participants}
A total of 24 participants (12 females and 12 males) were recruited for this study through online platforms and recruitment poster advertisements in local communities. Their average age was 25.2 years (SD = 3.71, range = 19–34). They received a study overview, gave informed consent, and were compensated about \$20 USD. \CHG{This study was approved by our institution's Research Ethics Board (REB).} Twenty-one were non-native English speakers, all with postsecondary education experience in English-speaking countries. On average, participants reported 6.5 hours of daily screen reading.

\subsection{Study Design}
This study employed a within-subjects design. Participants completed two tasks, namely a reading task (Task-\readText{\textbf{Read}}) and then followed by a matching task (Task-\matchText{\textbf{Match}}).
To control order effects, all conditions within each task were fully counterbalanced across 24 permutations, with each participant assigned a unique order. 

\subsection{Assessing Text Legibility on Draped UIs (Task-\readText{\textbf{Read}})}

Reading tasks must be carefully designed to reflect common reading patterns. Therefore, we used a task similar to that in Jankowski et al.~\cite{jankowski2010integrating}, which has been widely applied in prior studies \cite{serrano2016investigating, bernard2003comparing, darroch2005effect}. This task includes word substitution errors and requires participants to identify confusing words in a text passage.

\subsubsection{Passage Selection and Error Distribution}

To ensure consistent reading difficulty, all passages were drawn from a standardized English test corpus with controlled vocabulary~\cite{Quinn2007}, which has also been used in prior studies~\cite{sultana2024exploring}. A total of 48 passages were prepared, each containing 3 to 5 word substitution errors, yielding 192 errors in total. Passages were displayed within 12 lines in the interface with an average length of 98 words. Only complete sentences were included. Substitution positions covered most of the interface area \CHG{(see Appendix E.1)}. The 48 passages and 192 errors were evenly distributed across 4 experimental UI conditions, and therefore each condition contained 12 passages and 48 errors.

\subsubsection{Passage-Interface Pairing and Text Presentation}
The order of passages was fixed for all participants to ensure consistent difficulty, while the 6 interface shapes were randomized so that each participant encountered each shape twice across the 12 passages in one UI condition.
We used the sans-serif font Arial, which is generally preferred for reading on computer screens~\cite{bernard2003comparing}. The font size was 24 dmm (distance-independent millimeter, suggested by Google~\cite{designing_screen_interfaces_vr_2017}), assuming a viewing distance of 50 cm, which is the recommended size for body or button text in VR~\cite{designing_screen_interfaces_vr_2017}. 

\subsubsection{Procedure}

Participants were instructed to read each passage once as accurately and quickly as possible and to verbally report any words that seemed irrelevant or incorrect. They were allowed to move their heads freely to facilitate recognition but were asked to avoid habitual body movements that could introduce noise into viewpoint measurements \cite{sechrest1971occurrence, oshio2018shake}. For each reported word, the researcher recorded whether it was a true substitution; incorrect or omitted responses were counted as errors. After finishing a passage, participants said “done” or “next,” and the next passage was presented. They were not informed of the number of substitutions per passage.
After completing 12 passages in a given condition, participants filled out questionnaires for that condition and then rested before proceeding to the next condition. Before the main experiment, they completed a practice session with passages in the \textit{Snug} and \textit{Flat} conditions. These practice passages were excluded from the main test. The main experiment began once participants confirmed that they understood the task and completed practice without mistakes.

\subsubsection{Measures}
Objective measures included average viewpoint movement in meters recorded with an HMD during reading, mean reading time in seconds for each passage, and reading performance evaluated using the F1 score multiplied by 100 to present the results on a 0-100 scale.
Subjective measures included task workload assessed with the NASA Task Load Index (NASA-TLX) on a 7-point Likert scale \cite{hart1988development} and usability assessed with the pragmatic quality subscale of the User Experience Questionnaire Short version (UEQ-S) on a 7-point Likert scale \cite{hinderks2017design}. The UEQ-S evaluates interaction qualities associated with the user’s goals or tasks.
Participants evaluated Readability on a 7-point Likert scale. In addition, participants provided qualitative feedback regarding character recognition and depth-related visual comfort, such as double vision or accommodation difficulties. An open-ended question was also included to gather further insights into factors that facilitated or hindered the task.

\subsection{Results and Discussion (Task-\readText{\textbf{Read}})}

\begin{figure*}[htbp]
    \centering
    \includegraphics[width=\textwidth]{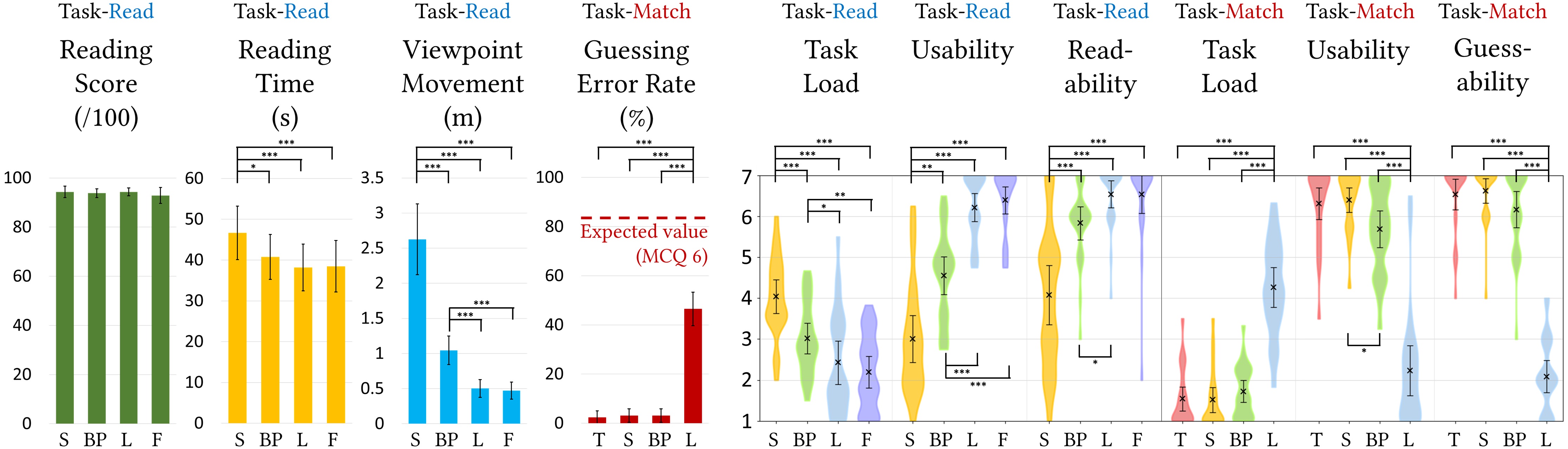}
    \caption{\CHG{Results of objective and subjective measures in Task-\readText{\textbf{Read}}/Task-\matchText{\textbf{Match}} of User Study II. MCQ 6 in the Guessing Error Rate refers to the error rate that occurs by chance when answering 6 multiple-choice questions (MCQ) randomly. T, S, BP, L, and F represent \textit{Tight}, \textit{Snug}, \textit{Balanced Point}, \textit{Loose}, and \textit{Flat} UI conditions, respectively. The error bars indicate the 95\% confidence interval. Asterisks indicate Bonferroni-corrected $p_c$ values: $p_c$ < .05 (*), $p_c$ < .01 (**), $p_c$ < .001 (***). Detailed descriptive statistics are provided in Appendix E.2.}}
    \Description{This figure presents the objective and subjective results from the reading task and the matching task in User Study Two. For the guessing-error measurement, the value called MCQ six represents the error rate that would occur purely by chance if a person were to guess randomly on a six-option multiple-choice question. The figure compares five interface conditions. These conditions are named Tight, Snug, Balanced Point, Loose, and Flat. The vertical error bars on each graph show the ninety-five percent confidence interval of the measurements. Asterisks mark the results that remain statistically significant after applying the Bonferroni correction. One asterisk indicates significance below point zero five, two indicate significance below point zero one, and three indicate significance below point zero zero one. Additional descriptive information for all measures can be found in Appendix E point two.}
    \label{fig:study1_results}
\end{figure*}

We used the Friedman test for all measures of Task-\readText{\textbf{Read}} (Table~\ref{tab:friedman_all}), as the data did not conform to normality. All results for Task-\readText{\textbf{Read}} and post-hoc pairwise comparisons using Wilcoxon signed-rank tests with Bonferroni-corrected $p_c$ values are reported in Fig.~\ref{fig:study1_results} and Table~\ref{tab:pairwise1}, respectively.

\begin{table}[h]
    \centering
    \caption{Friedman test results presented as ($\chi^2(3)$, $p$, $W$) for each measure in Task-\readText{\textbf{Read}} and Task-\matchText{\textbf{Match}} in User Study II.}
    \resizebox{\columnwidth}{!}{%
    \begin{tabular}{lcc}
        \toprule
        Measures & Task-\readText{\textbf{Read}} & Task-\matchText{\textbf{Match}} \\
        \midrule
        Reading Score (/100)    & (1.65, $\;\,$.6483,$\,$ .023) & -- \\
        Reading Time (s)        & (26.6, <0.001, .369)   & -- \\
        Viewpoint Movement (m)  & (66.6, <0.001, .925)   & -- \\
        Guessing Error Rate (\%)         & --                      & (59.57, <0.001, .827) \\
        Task Load (1--7)        & (50.0, <0.001, .695)   & (52.00, <0.001, .722) \\
        Usability (1--7)        & (60.0, <0.001, .834)   & (51.54, <0.001, .716) \\
        Readability (1--7)      & (48.2, <0.001, .669)   & -- \\
        Guessability (1--7)     & --                      & (54.88, <0.001, .762) \\
        \bottomrule
    \end{tabular}
    }
    \label{tab:friedman_all}
\end{table}

\begin{table*}[h]
\centering
\caption{Pairwise Wilcoxon signed-rank test results for condition comparisons in Task-\readText{\textbf{Read}}, reported as ($|Z|$, $p_c$, $|r|$). Metrics are in the order: Reading Score, Reading Time, Viewpoint Movement, Task Load, Usability, Readability. Reading Score column was omitted with ``-'' since the Friedman test was not significant. $^{*}$ for $p_c<0.05$, $^{**}$ for $p_c<0.01$, $^{***}$ for $p_c<0.001$. Notation $a\_b$ represents $a \times 10^{-b}$; e.g., $7.15\_7 = 7.15 \times 10^{-7}$}
\resizebox{\textwidth}{!}{%
\begin{tabular}{lcccccc}
\toprule
Comparison & Reading Score & Reading Time & Viewpoint Movement & Task Load & Usability & Readability \\
\midrule
\textit{Snug} vs. \textit{B.P.} & - & (2.900, *{{\scriptsize $(1.51\_2)$}}, .5920) & (4.271, ***{{\scriptsize $(7.15\_7)$}}, .8719)  & (4.243, ***{{\scriptsize $(1.43\_6)$}}, .8661) & (3.452, **{{\scriptsize $(3.27\_3)$}}, .7198) & (3.901, ***{{\scriptsize $(4.44\_4)$}}, .8723)  \\
\textit{Snug} vs. \textit{Loose} & - & (3.814, ***{{\scriptsize $(1.21\_4)$}}, .7786) & (4.271, ***{{\scriptsize $(7.15\_7)$}}, .8719) & (4.186, ***{{\scriptsize $(3.58\_6)$}}, .8544) & (4.271, ***{{\scriptsize $(7.15\_7)$}}, .8719) & (3.997, ***{{\scriptsize $(3.56\_4)$}}, .8722)  \\
\textit{Snug} vs. \textit{Flat} & - & (4.014, ***{{\scriptsize $(2.36\_5)$}}, .8194) & (4.271, ***{{\scriptsize $(7.15\_7)$}}, .8719)  & (4.271, ***{{\scriptsize $(7.15\_7)$}}, .8719) & (4.271, ***{{\scriptsize $(7.15\_7)$}}, .8719) & (3.893, ***{{\scriptsize $(5.47\_4)$}}, .8495)  \\
\textit{B.P.} vs. \textit{Loose} & - & (1.843, .3879, .3762) & (4.271, ***{{\scriptsize $(7.15\_7)$}}, .8719)  & (2.781, *{{\scriptsize $(3.23\_2)$}}, .6219) & (4.271, ***{{\scriptsize $(7.15\_7)$}}, .8719) & (2.669, *{{\scriptsize $(3.37\_2)$}}, .6892) \\
\textit{B.P.} vs. \textit{Flat} & - & (1.757, .4724, .3587) & (4.271, ***{{\scriptsize $(7.15\_7)$}}, .8719)   & (3.376, **{{\scriptsize $(4.36\_3)$}},.7199) & (3.969, ***{{\scriptsize $(4.24\_4)$}}, .8276) & (2.225, .1350, .5396) \\
\textit{Loose} vs. \textit{Flat} & - & (0.014, 1.000, .0029) & (1.843, .3879, .3762)  & (1.422, .9244, .3555) & (1.163, 1.000, .2909) & (0.000, 1.000, .0000) \\
\bottomrule
\end{tabular}%
}
\label{tab:pairwise1}
\end{table*}

For legibility assessed with objective measures, Reading Score did not differ significantly across any of the UI conditions, including the \textit{Flat} baseline. This implies that even in the highly curved \textit{Snug} condition, participants can read the text. Other measures, however, showed clear differences. Reading Time was significantly longer in the \textit{Snug} condition than in all others, with no \CHG{significant} differences among the remaining conditions. Viewpoint Movement, derived from HMD tracking, was likewise largest in the \textit{Snug} condition. The \textit{B.P.} condition required more head movement than \textit{Loose} and \textit{Flat}, but less than \textit{Snug}.

In subjective measures, Task Load was highest in the \textit{Snug} condition, with \textit{B.P.} rated higher than \textit{Loose} and \textit{Flat}, but lower than \textit{Snug}. Usability for reading, along with direct ratings of readability, was lowest in the \textit{Snug} condition. Although \textit{B.P.} was rated lower than \textit{Flat} and \textit{Loose}, it was significantly better than \textit{Snug} in both usability and readability. Taken together, these findings suggest that despite its curved shape, the \textit{B.P.} condition mitigated reading difficulties and maintained readability comparable to that of a flat display.

In the additional qualitative feedback, 22 participants reported no character recognition problems across all conditions. Only 2 participants (\textasciitilde8.3\%) noted difficulty recognizing some words on the edge in the \textit{Snug} condition. Along with the objective Reading Score, this indicates that further smoothing beyond \textit{B.P.} might not cause recognition difficulties severe enough to hinder reading performance for most users.

Moreover, all participants reported no focal length issues in any condition, suggesting that our method may be robust against accommodation and double vision in reading tasks.

\begin{table*}[ht]
\centering
\caption{Pairwise Wilcoxon signed-rank test results for condition comparisons in Task-\matchText{\textbf{Match}}, reported as ($|Z|$, $p_c$, $|r|$). Metrics are in the order: Guessing Error Rate, Task Load, Usability, Guessability. $^{*}$ for $p_c<0.05$, $^{**}$ for $p_c<0.01$, $^{***}$ for $p_c<0.001$. Notation $a\_b$ represents $a \times 10^{-b}$; e.g., $7.15\_7 = 7.15 \times 10^{-7}$}
\resizebox{0.8\textwidth}{!}{%
\begin{tabular}{lcccc}
\toprule
Comparison & Guessing Error Rate & Task Load & Usability & Guessability \\
\midrule
\textit{Tight} vs. \textit{Snug} & (0.338, 1.000, .1278)   & (0.282, 1.000, .0755) & (0.170, 1.000, .0440) & (0.222, 1.000, .0670) \\
\textit{Tight} vs. \textit{B.P.} & (0.280, 1.000, .0990)   & (1.325, 1.000, .3215) & (2.243, .1450, .5286) & (1.188, 1.000, .3295) \\
\textit{Tight} vs. \textit{Loose} & (4.271, ***{{\scriptsize $(7.15\_7)$}}, .8719)  & (4.271, ***{{\scriptsize $(7.15\_7)$}}, .8719) & (4.271, ***{{\scriptsize $(7.15\_7)$}}, .8719) & (4.271, ***{{\scriptsize $(7.15\_7)$}}, .8719) \\
\textit{Snug} vs. \textit{B.P.}  & (0.105, 1.000, .0428)   & (1.775, .4523, .4305) & (2.757, *{{\scriptsize $(3.42\_2)$}}, .6324) & (2.236, .0830, .6202) \\
\textit{Snug} vs. \textit{Loose} & (4.271, ***{{\scriptsize $(7.15\_7)$}}, .8719)   & (4.271, ***{{\scriptsize $(7.15\_7)$}}, .8719) & (4.271, ***{{\scriptsize $(7.15\_7)$}}, .8719) & (4.271, ***{{\scriptsize $(7.15\_7)$}}, .8719) \\
\textit{B.P.} vs. \textit{Loose} & (4.271, ***{{\scriptsize $(7.15\_7)$}}, .8719)   & (4.271, ***{{\scriptsize $(7.15\_7)$}}, .8719) & (4.243, ***{\scriptsize $(1.43\_6)$}, .8661) & (4.271, ***{\scriptsize $(7.15\_7)$}, .8719) \\
\bottomrule
\end{tabular}%
}
\label{tab:pairwise2}
\end{table*}

\subsection{Assessing \CHG{Shape Awareness} (Task-\matchText{\textbf{Match}})}
The matching task used the same 6 shapes as the reading task. Two different interface textures were employed in the experiments, both simulated computer windows but had different layouts, each containing a combination of figures, words, and images (Fig.~\ref{fig:test_environments}(c)). By pairing the two textures with the 6 shapes, a total of 12 combinations were generated per condition. Participants experienced each shape twice in a randomized order.

\subsubsection{Procedure}

Before the main task, participants completed a practice session in a virtual office containing the 6 objects. They placed the interface over each object and manually adjusted the iteration level to observe shape changes using a mouse and keyboard. A white gray checkered pattern (1 cm × 1 cm, color code \#7F7F7F) was used during practice to make deformation and shading more visible. After all objects and deformation behaviors were experienced, the main experiment began.
All participants first saw each shape at its actual scale. During the real test, the corner shape was scaled down to fit the multiple choice layout on the desk due to space limitations (Fig.~\ref{fig:test_environments}(b) right). For the main task, participants completed trials in a unique sequence of Draped UI conditions assigned to them.
In each trial, an interface shape was shown and participants selected from 6 options the object they believed had generated it. All 12 texture-shape combinations were presented in random order. After 12 trials for a condition, participants removed the HMD, completed a questionnaire for that condition, and took a break before continuing.

\subsubsection{Measures}

As an objective measure of matching, error rates were recorded for each condition by dividing the number of errors by 12 and multiplying by 100 to convert them into percentage values. Task workload and usability were measured as in the reading task using the NASA-TLX and the pragmatic quality subscale of the UEQ-S, both on 7-point Likert scales. To assess subjective matching, defined as the perceived ability to infer the object that generated an interface from its shape, participants answered the customized question \textit{``In this condition, how well can you guess the object behind the UI?''} They also responded to the qualitative question \textit{``What made this task easier or more difficult?''} to explore factors affecting task difficulty. 

\subsection{Results and Discussion (Task-\matchText{\textbf{Match}})}

We used the Friedman test for all measures of Task-\matchText{\textbf{Match}} (Table~\ref{tab:friedman_all}), as the data did not conform to normality. All results for Task-\matchText{\textbf{Match}} and post-hoc pairwise comparisons using Wilcoxon signed-rank tests with Bonferroni-corrected $p_c$ values are reported in Fig.~\ref{fig:study1_results} and Table~\ref{tab:pairwise2}, respectively.

The objective metric, the Guessing Error Rate, showed that the \textit{B.P.} condition performed on par with the tighter configurations (\textit{Tight}, \textit{Snug}), while the \textit{Loose} condition produced markedly elevated errors. This demonstrates that \textit{B.P.} preserved recognition accuracy without the severe breakdown observed in \textit{Loose}. 
For the subjective measures, Task Load under \textit{B.P.} was significantly lower than in \textit{Loose} but higher than in \textit{Tight} and \textit{Snug}; usability and guessability followed the same pattern, with \textit{B.P.} consistently differing from \textit{Loose}. However, in terms of usability, \textit{B.P.} scored significantly lower than \textit{Snug}. Importantly, prior findings in Task-\readText{\textbf{Read}} suggested that tighter configurations (i.e., \textit{Snug}) impair readability performance. In contrast, \textit{B.P.} supports legibility while also avoiding the high error rates of \textit{Loose}, thereby achieving a balanced compromise.

Additional Wilcoxon tests confirmed that all conditions performed below the chance error rate of 83.33\% ($|Z|= 4.2857, p < .001, |r|=0.8748$), even under the \textit{Loose} condition where shape discrimination was most challenging, indicating that the results were not due to chance.
Taken together, these findings suggest that \textit{B.P.} achieves a balanced performance, avoiding the shape awareness collapse, and supporting legible surface representation.

\subsection{User Study II: Limitations}

\subsubsection{Task-\readText{\textbf{Read}}}
This study explored reading performance on arbitrary 3D surfaces, though it does not cover all possible factors. We generated interfaces for typical objects found in everyday office environments and performed reading tasks, showing that even under complex surface conditions the Reading Score was comparable to flat displays. However, these results cannot be assumed to hold universally, as outcomes vary depending on how characters are rendered, or how the objects are shaped. For instance, typographic variables such as font, interline spacing, and margin sizes were not systematically tested, and situations occurred where small objects are overshadowed by surrounding larger ones. \CHG{Also, participants language proficiency differences were not formally controlled, which could affect reading speed and strategies such as skimming, scanning, or using background knowledge \cite{hayashi1999reading}. Future studies could include standardized proficiency measures to address this limitation and to better understand the impact of language skills on legibility.}

Additional insights from user feedback highlight reading challenges on arbitrary 3D surfaces that go beyond character recognition, such as locating the next line (P05, P07, P18), dealing with unfamiliar content (P09), recognizing edge characters (P12), and finding an optimal viewpoint (P24). These observations reveal novel tasks in reading on irregular shapes and suggest opportunities for extending interface design. Moreover, while our evaluation focused on characters, we did not assess the interpretability of symbols or graphs that often appear in practical settings. This points toward a future research direction examining broader information types beyond text. Also, the study was conducted within a laboratory setting rather than in situ, which constitutes a limitation and suggests the need for future work in diverse real-world reading contexts.
Our results suggest that the \textit{B.P.} condition makes the interface legible without significantly reducing reading performance, but such effects may still be achieved through other iterations. \CHG{Rather, these alternatives could provide users with more choices}, highlighting the need for broader investigation. Furthermore, our implementation relies on a relatively naive topology and mapping, suggesting that distortion could be further optimized through surface parameterization~\cite{floater2005surface} and remeshing~\cite{hormann2001remeshing}. \CHG{Also, our study prioritized understanding how real users read in everyday contexts rather than applying computational approaches to legibility. Future work could incorporate techniques such as Optical Character Recognition (OCR) to assess legibility and support more detailed, quantitative analyses that enhance interface design.}

\subsubsection{Task-\matchText{\textbf{Match}}}
We examined the guessability of surrounding objects, with the important assumption that this process depends on participants’ preknowledge. Our method provided partial shape information on the interface surface, enabling inference of the underlying object. However, when such cues are insufficient for distinguishing objects regardless of their size or form, users may fail to make accurate guesses. For example, one participant (P07) noted, “If shapes look similar to each other, it is not easy to distinguish,” underscoring the limitations of inference when shape cues are minimal.
This shows that inference is feasible when users already know what to expect in familiar environments, but its reliability decreases under unfamiliar or ambiguous conditions, thus requiring further research. For example, using spatial memory to support understanding of the surrounding environment~\cite{hurter2024memory} can be considered as one aspect of such future research questions.
Furthermore, our study tested only two interface layouts related to shape, leaving broader design variations such as color and visual patterns that may influence user perception unexamined. Future research can include systematic comparisons across a wider range of layouts.

\section{Extended Interaction Opportunities Enabled by \CHG{CAMEO}}\label{sec:discussions}

\begin{figure}[htbp]
    \centering
    \includegraphics[width=\columnwidth]{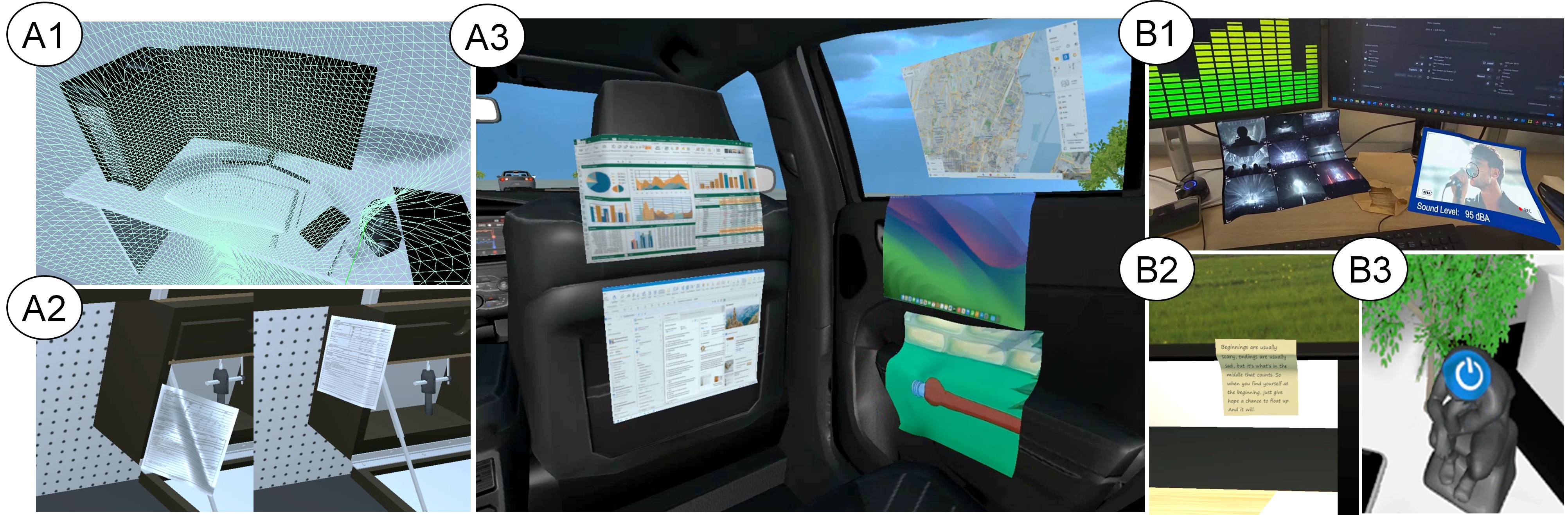}
    \caption{\CHG{Design opportunities emerging from CAMEO. A1: Contour-based continuous content flow across surfaces and space. A2: Surface-adaptive panning across real-world geometries. A3: Seamless CAMEO workspace continuity over angular structures. B1: Surface-based minimization of UI windows near the user. B2: Micro UIs represented as sticky notes attached to relevant real-world objects. B3: Object-anchored adaptive UI buttons conforming to physical geometries.}}
    \Description{Design opportunities emerging from CAMEO. A1: Contour-based continuous content flow across surfaces and space. A continuous CAMEO-defined spatial surface is depicted near a desktop, showing content flowing smoothly across nearby physical surfaces. A2: Surface-adaptive panning across real-world geometries. The illustration shows an interface panning across connected surfaces through CAMEO, adapting naturally to real-world shapes. A3: Seamless CAMEO workspace continuity over angular structures. This example shows how CAMEO can operate inside a vehicle, with multiple interfaces adaptively arranged along nearby surfaces. B1: Surface-based minimization of UI windows near the user. In a desktop setting, interface windows are draped and minimized beneath the monitor along the physical surface. B2: Micro UIs represented as sticky notes attached to relevant real-world objects. Sticky-note–like micro interfaces appear beneath the monitor, attached to the related physical surface. B3: Object-anchored adaptive UI buttons conforming to physical geometries. On the desk, an object displays a button icon placed through CAMEO, conforming to the object’s physical geometry.}
    \label{fig:discussion}
\end{figure}

While this work focused on legibility and background awareness, the core implementation of \CHG{CAMEO} opens up broader opportunities for interaction and workspace management. Because \CHG{CAMEO} flexibly conforms to immediate surfaces, it can be readily adapted to support new layout strategies that extend MR workspaces beyond conventional floating windows. With minor modifications, \CHG{CAMEO} enables novel interaction scenarios that make direct use of surrounding physical surfaces as part of the interface design. Below, we illustrate opportunities for future implementations across two key categories: \textit{Seamless UI Flow} and \textit{Semantic Positioning}. 

\subsection{\CHG{Seamless Flow Across CAMEO Surfaces}}

\CHG{CAMEO’s smooth and seamless surface connections overcomes complex environmental geometry and extends the workspace into the ambient flow of the real world.}

\subsubsection{Continuous Transitions Across Surfaces and Space}

Building on \CHG{CAMEO}’s near-surface approach, seamless transitions can be supported not only between on-surface and mid-air interactions but also across multiple surfaces. By leveraging continuous contour-based representations \CHG{(Fig.~\ref{fig:discussion}(A1))}, \CHG{CAMEO} allows users to fluidly reposition and manipulate information across intermediate spaces, enabling layout strategies that do not disrupt task flow. Beyond continuity, this approach introduces new spatial metaphors: content can “flow” across the workspace like a ribbon or “bridge” between surfaces to maintain visual and functional coherence. Such transitions transform surrounding geometry into an active medium for interaction, allowing users to think of their workspace as a continuous spatial fabric rather than a set of isolated screens.

\subsubsection{Expanded Panning on Surfaces}

\CHG{CAMEO} supports panning actions that extend content across physical surfaces such as desks or walls, or even over complex geometries \CHG{(Fig.~\ref{fig:discussion}(A2))}. For example, panning from a smartphone into the environment allows users to view broader content regions while preserving functional awareness of their surroundings. More than simply enlarging display space, this approach enables hybrid layouts in which digital content can be anchored to both device screens and the contours of nearby surfaces. As users pan, information can dynamically adapt to the surface geometry, unfolding like a scroll across walls, desks, or objects. This transforms everyday environments into continuous canvases, where content can scale fluidly from pocket-sized devices to room-scale spatial layouts.

\subsubsection{Continuity Across Angular Spaces}

By treating all surfaces as part of a single continuous shared workspace, \CHG{CAMEO} can support fluid layouts even when those surfaces meet at orthogonal angles \CHG{(Fig.~\ref{fig:discussion}(A3))}. For example, in a car interior, a \CHG{CAMEO} workspace can drape seamlessly from the seat cushion onto the adjacent door panel, maintaining a coherent interface across two sharply perpendicular surfaces. Whereas planar MR windows would fragment or require repositioning at such corners, \CHG{CAMEO} preserves both continuity and legibility as it adapts to the underlying geometry. This enables users to view or manipulate information across the seat–door boundary as if it existed on one unified surface, demonstrating how the approach accommodates complex real-world structures without interrupting interaction.

\subsection{\CHG{Positioning UIs with Semantics}}

\CHG{CAMEO creates a conceptual flow between continuous virtual surfaces and real-world objects. By leveraging users’ familiarity with the physical environment, CAMEO can bind digital content to real-world structures, enriching interaction through contextually meaningful spatial integration.}

\subsubsection{Surface-Based Minimization}

\CHG{CAMEO} can be leveraged to reimagine the conventional minimization function by anchoring windows onto nearby surfaces \CHG{(Fig.~\ref{fig:discussion}(B1))} rather than collapsing them into a taskbar. The user can quickly transform the UI from \CHG{CAMEO} to a flat surface by gaze or other interactive method. Minimized content could remain visible along the edge of a desk or wall, preserving quick access while freeing space for other tasks. Beyond saving space, \CHG{CAMEO} enables spatially meaningful positioning, where windows can be tucked into semantically relevant physical areas. For example, a chat window minimized onto the rim of a coffee mug as a gentle notification, a sticky note at the edge of the monitor \CHG{(Fig.~\ref{fig:discussion}(B2))} or a reference document collapsed along the edge of a notebook. These anchors are not only persistent but can be designed to be context-aware, resurfacing when users gaze or approach the corresponding object or surface. This transforms physical surroundings into a dynamic memory scaffold for multitasking, blending recall, context, and quick access in ways not achievable with flat MR interfaces.

\subsubsection{Adaptive Buttons Anchored to Objects}

Through its placement flexibility, \CHG{CAMEO} enables UI elements like buttons or icons to be directly attached to nearby physical artifacts \CHG{(Fig.~\ref{fig:discussion}(B3))}. For instance, a control panel button can be embedded on each device in a workspace, creating intuitive and spatially meaningful mappings between digital functions and physical objects. Unlike flat mid-air interfaces, these anchored buttons can contour around an object’s surface, enabling affordances that leverage its geometry (e.g., a volume slider wrapping around a speaker’s dial or a progress bar following the edge of a laptop). Unlike widgets that directly get placed on objects, this integration transforms objects into hybrid interactive devices, where their physical form directly guides and constrains digital controls.

\section{Conclusion}\label{sec:conclusion}

\CHG{This work introduced Contour-Adaptive Mixed Environment Overlays (CAMEO), a novel contour-aware MR workspace that integrates digital content with the user’s physical surroundings. By draping virtual windows over real-world surfaces, CAMEO addresses the trade-off between immersive digital interaction and sustained awareness of nearby objects. Our empirical studies demonstrated that CAMEO not only reduces interaction inefficiencies, such as hand-movement detours, but also preserves text legibility across complex surfaces, positioning it as a viable complement to conventional flat, mid-air MR interfaces. Beyond its technical validation, CAMEO establishes a new design paradigm for MR workspaces that harmonizes immersion, legibility, and environmental understanding, and offers meaningful implications for further inquiry. Finally, we believe MR interfaces should evolve toward more natural integration with the physical world. Our approach supports this by enabling virtual workspaces to blend seamlessly into real-world contexts.}

\begin{acks}
We thank the participants for their time and cooperation. We are also grateful to the members of our laboratory for their valuable feedback and support. Our thanks also go to the anonymous reviewers for their constructive comments. Finally, this work was supported by the Natural Sciences and Engineering Research Council of Canada (NSERC).
\end{acks}

\bibliographystyle{ACM-Reference-Format}
\bibliography{References/references}

\end{document}